\documentclass[manuscript]{acmart}
\usepackage{bbding} 
\usepackage{epigraph}
\usepackage{hyperref}
\usepackage{booktabs}

\usepackage{amsmath}

\newtheorem{hyp}{Hypothesis}

\newtheorem{obs}{Observation}

\input{definitions}

\AtBeginDocument{%
  \providecommand\BibTeX{{%
    \normalfont B\kern-0.5em{\scshape i\kern-0.25em b}\kern-0.8em\TeX}}}

\setcopyright{acmcopyright}
\copyrightyear{2022}
\acmYear{2022}
\acmDOI{}

\acmConference[]{}{}{}
\acmBooktitle{}
\begin{document}

% \title{Counterfactual Interactive Recommender System via Debiasing Offline Reinforcement learning}
\title{CIRS: Bursting Filter Bubbles by Counterfactual Interactive Recommender System}
% \subtitle{An offline reinforcement learning framework with debiased user model.}

% \author{
%     Chongming Gao$^{1}$,
%     Wenqiang Lei$^{2}$, 
%     Jiawei Chen$^{1}$,
%     Shiqi Wang$^{3}$,
%     Xiangnan He$^{1,*}$,\\
%     Shijun Li$^{1}$, 
%     Biao Li$^{4}$, 
%     Yuan Zhang$^{4}$,
%     Peng Jiang$^{4}$
%     }

% \thanks{
%     $^*$ Corresponding author. 
% }
% \affiliation{
%     $^{1}$University of Science and Technology of China; 
%     $^{2}$Sichuan University, China; \\
%     $^{3}$Chongqing University, China;
%     $^{4}$Kuaishou Technology Co., Ltd.\\
%     chongming.gao@gmail.com, 
%     wenqianglei@gmail.com,
%     cjwustc@ustc.edu.cn,
%     shiqi@cqu.edu.cn,
%     xiangnanhe@gmail.com,
%     lishijun@mail.ustc.edu.cn, 
%     biaoli6@139.com,
%     yuanz.pku@gmail.com,
%     jp2006@139.com,
%     \country{}
% }

\author{Chongming Gao}
\affiliation{%
  \institution{University of Science and Technology of China}
  % \streetaddress{1 Th{\o}rv{\"a}ld Circle}
  \city{Hefei}
  \country{China}}
\email{chongming.gao@gmail.com}

\author{Shiqi Wang}
\affiliation{%
  \institution{Chongqing University}
  \city{Chongqing}
  \country{China}}

\author{Shijun Li}
\affiliation{%
  \institution{University of Science and Technology of China}
  % \streetaddress{1 Th{\o}rv{\"a}ld Circle}
  \city{Hefei}
  \country{China}}

\author{Jiawei Chen$^*$}
\affiliation{%
  \institution{Zhejiang University}
  % \streetaddress{1 Th{\o}rv{\"a}ld Circle}
  \city{Hangzhou}
  \country{China}}

\author{Xiangnan He$^*$}
\affiliation{%
  \institution{University of Science and Technology of China}
  % \streetaddress{1 Th{\o}rv{\"a}ld Circle}
  \city{Hefei}
  \country{China}}
\email{xiangnanhe@gmail.com}

\author{Wenqiang Lei}
\affiliation{%
  \institution{Sichuan University}
  \city{Chengdu}
  \country{China}
}

\author{Biao Li}
\affiliation{%
  \institution{Kuaishou Technology Co., Ltd.}
  \city{Beijing}
  \country{China}}

\author{Yuan Zhang}
\affiliation{%
  \institution{Kuaishou Technology Co., Ltd.}
  \city{Beijing}
  \country{China}}

\author{Peng Jiang}
\affiliation{%
  \institution{Kuaishou Technology Co., Ltd.}
  \city{Beijing}
  \country{China}}

\thanks{
    $^*$ Corresponding author. 
}

\renewcommand{\shortauthors}{Gao, et al.}

\begin{abstract}

While personalization increases the utility of recommender systems, it also brings the issue of \textit{filter bubbles}. E.g., if the system keeps exposing and recommending the items that the user is interested in, it may also make the user feel bored and less satisfied. 
Existing work studies filter bubbles in static recommendation, where the effect of overexposure is hard to capture. In contrast, we believe it is more meaningful to study the issue in interactive recommendation and optimize long-term user satisfaction.
Nevertheless, it is unrealistic to train the model online due to the high cost. As such, we have to leverage offline training data and disentangle the causal effect on user satisfaction. 
% The main challenge lies in: since we cannot train the model online, we have to disentangle the causal effect on user satisfaction on the offline training data.

To achieve this goal, we propose a counterfactual interactive recommender system (CIRS) that augments offline reinforcement learning (offline RL) with causal inference. The basic idea is to first learn a causal user model on historical data to capture the overexposure effect of items on user satisfaction. It then uses the learned causal user model to help the planning of the RL policy. 
% By leveraging the causal model, the policy can disentangle the intrinsic user preference from the overexposure effect of items. Thus, we can tackle the filter bubble problem during optimizing cumulative user satisfaction. 
To conduct evaluation offline, we innovatively create an authentic RL environment (KuaiEnv) based on a real-world fully observed user rating dataset. 
% In this environment, users will quit the interaction if filter bubbles emerge. 
The experiments show the effectiveness of CIRS in bursting filter bubbles and achieving long-term success in interactive recommendation. The implementation of CIRS is available via \textcolor{magenta}{\url{https://github.com/chongminggao/CIRS-codes}}.

% To avoid the phenomenon and burst filter bubbles, it is necessary to disentangle the intrinsic user interest from the overexposure effect of items. 

\end{abstract}




\maketitle

\section{Introduction}
\label{sec:intro}

Recommender systems have deeply affected our lives. They change the way of retrieving information from searching strenuously, to obtaining conveniently via the precise personalization. The system usually achieves personalization by learning on the collected behavior data of users and selecting the products that users potentially like \citep{MFmodel,DeepFM,lightgcn,gao2019bloma}. With time evolving and data accumulating, the recommender gradually becomes a mirror reflecting each user's interest and narrows down the recommendation lists to the items with the maximum user interest. 
However, this comes at a price. While enjoying the precise personalized recommendations, users have to face the fact that the variety of information shrinks. When users become isolated from the information that varies from their dominant preferences, they are stuck in the so-called ``filter bubbles''.

Filter bubbles are common in recommender systems. Recent studies conducted extensive experiments in large-scale recommender systems and found there are two primary causes of filter bubbles \cite{WWW21fb,fb2014,Measuring_misinfo,how_youtube,Auditing_youtube,recsys21best}. From users' perspective, those who have less diverse preferences are more liable to be stuck in the bubbles. From the system's perspective, learning can lead to emphasizing the dominant interest of a user.
Moreover, the system usually assumes that user satisfaction equals intrinsic interest --- even though the items that user love have been overexposed, it assumes that the user satisfaction remains unchanged, which is improper. We believe that the over-exploitation behavior of the recommender algorithm is the main cause of filter bubbles, and we will use the overexposure effect as the proxy for evaluating filter bubbles.
% Consequently, apart from users’ own responsibility, filter bubbles exist in recommendations because the model holds the stereotype that users have unchanged satisfaction even items have been overexposed to them. In essence, the root cause of filter bubbles is that the model overexposes certain items to users. 

Overexposing items has a pernicious effect on a user's satisfaction, even though the user is interested in the recommended item. For example, a user who likes dancing can be satisfied when she receives a recommendation about a dancing song. However, she may be bored after dozens of times of incessant recommendations about dancing songs and thus refuses to choose it. Therefore, it is of great significance to burst filter bubbles for a recommender system for the purpose of maximizing user satisfaction. The key is to disentangle the causal effects on user satisfaction, i.e., modeling how a user's intrinsic interest together with the overexposure effect affect the user's final satisfaction.
 % (\myfig{intrograph}).

% In this paper, we only focus on the filter bubble caused by the recommendation algorithm.
% Filter bubbles have a pernicious effect on user experience, and we will verify this in the field study conducted on the short video social platform Kuaishou App\footnote{\url{https://www.kuaishou.com/}} (\mysec{verification}). 

% In other words, the model cannot follow the dynamic preference of users.

% In essence, filter bubbles results from the closed recommendation-feedback loop in the interaction recommendation process.

Existing methods address filter bubble with two strategies: (1) helping users improve their awareness of diverse social opinions \cite{fb_awreness,echo_1}, and (2) making the model enhance diversity \cite{liu2020diversified,Deconstructing_fb,echo_2}, serendipity \cite{combating_fb,neural_serendipity}, or fairness \cite{AAAI2020burstingfb} of the recommendation results. However, most of these strategies are heuristic and focus on the static recommendation setting.
% , where the effect of item overexposure is hard to capture and evaluate. In other words, these solutions
They cannot fundamentally address the problem since they do not directly consider the main cause of the filter bubble by modeling the overexposure effect.

\begin{figure}[!t]
\centering
\includegraphics[width=0.6\linewidth]{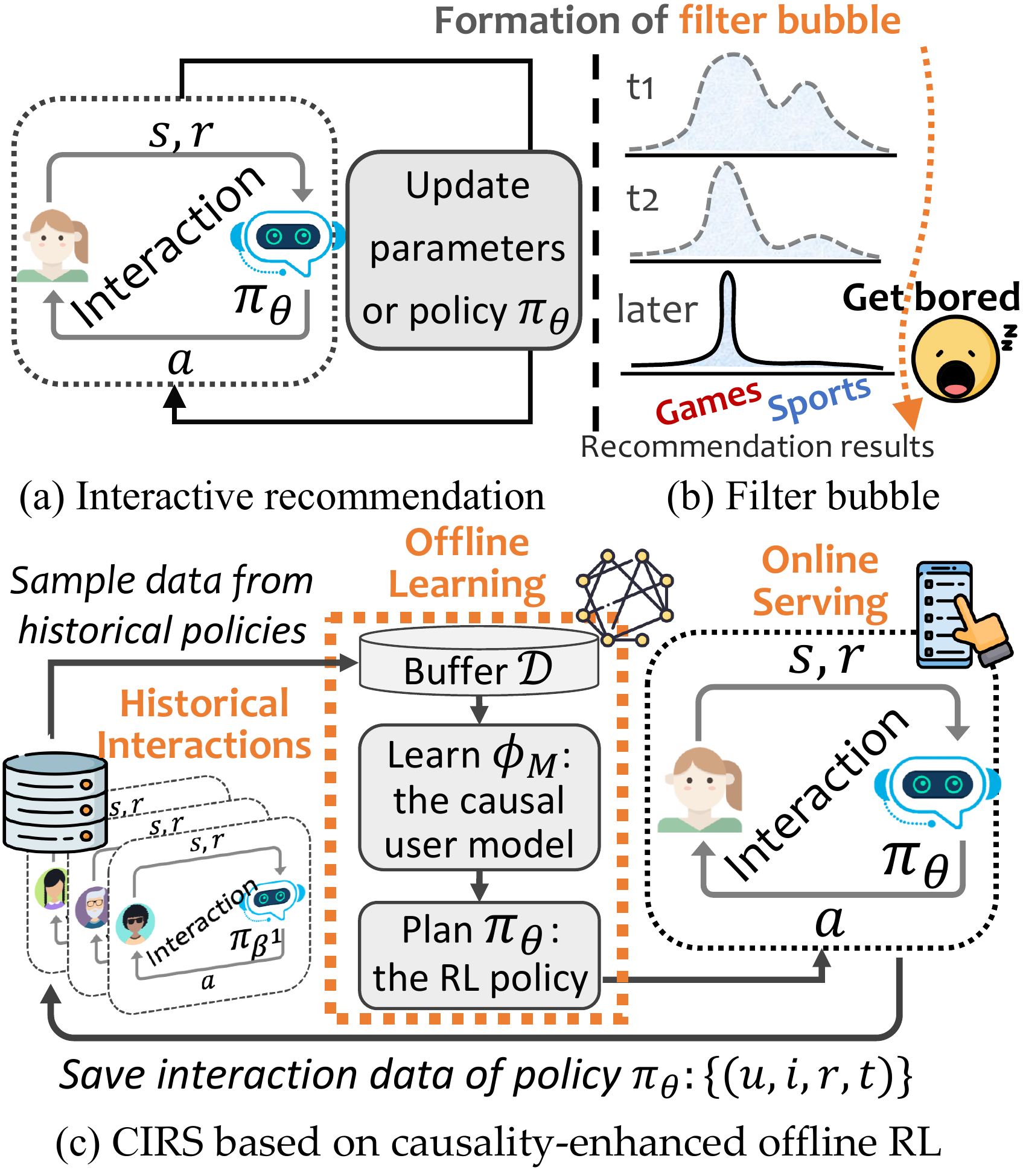}
% \vspace{-2mm}
\caption {Illustration of interactive recommendation and the formation of the filter bubble.}
\label{fig:intro}
\end{figure}

\medskip
In this work, we focus on the interactive recommender system (IRS) in the dynamic environment. An IRS is formulated as a sequential decision-making process, which allows the tracking and modeling of the dynamic and real-time overexposure effect.
% we can naturally capture and evaluate the effect of filter bubbles. To be more specific, we aim to maximize the cumulative user satisfaction over the whole interaction process, 
% , where the filter bubble evolves naturally in the dynamic recommendation-feedback loop and thus can be easily modeled and evaluated. 
\myfig{intro}(a) draws an example of interactive recommendation, where a model recommends items (i.e., makes an action $a$) to a user based on the interaction context (i.e., state $s$ reflecting the user's information and interaction history), then it receives user feedback (i.e., reward $r$ representing user satisfaction). The interaction process is repeated until the user quits. The model will update its policy $\pi_\theta$ with the goal to maximize the cumulative satisfaction over the whole interaction process. To achieve this goal, the IRS should avoid overexposing items because users will get bored and their satisfaction will drop in filter bubbles (\myfig{intro}(b)).

% we aim to maximize the cumulative user satisfaction over the whole interaction process. Since user satisfaction will drop in filter bubbles, we need to avoid repeatedly recommending similar items in the interaction process. The interactive model can achieve this by learning to adapt to users' timely feedback. 

% In static recommendation, such as matrix factorization (MF)-based model \cite{MFmodel}, DeepFM \cite{DeepFM}, and lightGCN \cite{lightgcn}, the policy is static and straightforward: just updating the model parameters once receiving user feedback. Interactive recommender systems (IRSs) \cite{nips_model_bai,chen2019generative,counterfactual_reward} usually resort to reinforcement learning (RL) that designs a policy $\pi_\theta$ pursuing the long-term success, such as the long-term user engagement \cite{liu2018deep,wang2021reinforcement}, cumulative financial gain \cite{rl4profit}, or maximum diversity \cite{liu2020diversified}. 

Though appealing, this idea faces an inevitable challenge:
% Regardless of the effectiveness of existing interactive recommender systems (IRSs) \cite{liu2018deep,sigir20_interactive,zhang2019reward,NICF,lixin_kdd,liu2020diversified}, it is impractical to train an IRS online with real users because:
it is difficult to learn an IRS online with real-time feedback. The reason is twofold:
(1) for the model, training the policy online increases the learning time and the deploying complexity \cite{xu2015infrastructure};
(2) for the user, interacting with a half-baked system can hurt satisfaction \cite{schnabel2018short,gao2021advances}. 
Hence, it is necessary to train the IRS offline on historical logs before serving users online. 
To this end, we need to conduct causal inference \cite{pearl2009causality} on the offline data to disentangle the causal effect of user interest and overexposure. This provides an opportunity to answer a counterfactual question in the serving phase: ``\emph{Would the user still be satisfied with the interested item if it had been overexposed?}'' --- If the answer is no, then the filter bubble will occur once recommending the item, 
so we should not recommend it.

\smallskip
We propose a counterfactual interactive recommender system (CIRS) to achieve this goal. It enhances offline reinforcement learning (offline RL) \cite{levine2020offline} with causal inference \cite{pearl2009causality}. 
% CIRS is an unbiased model that can disentangle the users' intrinsic preferences from the overexposure effect of item, thus it can learn to avoid filter bubbles in the interactive process.
 \myfig{intro}(c) shows its learning framework, which contains three recurrent steps:
1) learning a causal user model to capture both user interest and item overexposure effect,
2) using the learned user model to provide \emph{counterfactual satisfaction} (i.e., the rewards given by the causal user model instead of immediate users' feedback) for planning an RL policy,
and 3) evaluating the RL policy in terms of the cumulative user satisfaction in real environments.
% At last, the evaluated data can be collected into the database for the next learning process.

% To evaluate the proposed method, 
% we conduct experiment on the benchmark environment for evaluating RL-based recommender, VirtualTaobao \cite{virtualtaobao}. 
In addition, we propose an interactive recommendation environment, Kuai\-Env, for evaluating the problem. 
% Unlike the existing environments (e.g., VirtualTaobao \cite{virtualtaobao}) which simulate user behaviors, 
It is created based on the KuaiRec dataset that contains a fully filled user-item matrix (density: 99.6\%) collected from the Kuaishou App\footnote{\url{https://www.kuaishou.com/cn}} \cite{gao2022kuairec}. In this matrix, IRSs can be evaluated more faithfully since there are no missing values.
To effectively reflect the effect of filter bubbles, we further add a "feel bored then quit" exit mechanism in both VirtualTaobao and KuaiEnv to simulate the reactions of real users. 
% We measure the recommendation performance by users' cumulative satisfaction. It means a good policy should not only pursue the large satisfaction in each round but also maintain a long interaction. 
The experiments show that the proposed CIRS can capture the overexposure effect in offline learning and achieve a larger cumulative satisfaction compared with other baselines. 

% \smallskip
Our contributions are summarized as follows:
\begin{itemize}
    \item To the best of our knowledge, this is the first work to address filter bubbles in interactive recommendation, where filter bubbles can be more naturally observed and evaluated via modeling the overexposure effect on user satisfaction.
    \item We solve the problem by integrating causal inference into offline RL. We are the first to combine causal inference and offline RL in interactive recommender systems.
    \item We conduct empirical studies on Kuaishou App to verify that overexposing items indeed hurts user experience, and we demonstrate that the proposed method can burst filter bubbles and increase cumulative satisfaction.
\end{itemize}

% adjust its recommendation lists and update its policy. The feedback loop exists until users quit using the system. \myfig{intro} (a) draws an example of the interactive recommendation process. Where $a$ is the action, i.e., the recommendation, made by the model. $s$ is the state reflecting the interaction context, and $r$ is the reward in user feedback representing user satisfaction. 

% $\pi_\theta$ is the target model policy and $\pi_\beta$ is the behavior policy in historical interaction. As the interaction proceeds, the 

\section{Related Work}
\label{sec:related}

We briefly review the related works from three perspectives: filter bubbles, causality inference, and offline RL in recommendation.

\subsection{Filter Bubbles in Recommendation}

\setlength{\epigraphwidth}{.98\columnwidth}
\renewcommand{\textflush}{flushepinormal}
\epigraph{\textit{``Personalization filters serve up a kind of invisible autopropaganda, indoctrinating us with our own ideas, amplifying our desire for things that are familiar and leaving us oblivious to the dangers lurking in the dark territory of the unknown."} }
{{\footnotesize{\textit{––Eli Pariser, The Filter Bubble \cite{pariser2011filter}}}}}

The term filter bubble was coined by internet activist Eli Pariser \cite{pariser2011filter} and used for describing the state of intellectual isolation that results from the algorithm's personalization process. 
\citet{bruns2019filter} extensively discussed the causes and effects of filter bubbles. He claimed that the filter bubbles can be not real under certain situation, instead, it is only used as a scapegoat for avoiding addressing deep-rooted problems by blaming technology. He also discussed the filter bubble from a psychological aspect and described its political effects on society. 

\citet{mckay2022turn} investigated the filter bubble by interviewing 18 people recruited from a variety of sources about their experience of changing views on an issue that was important to them. They found an unexpected result that, instead of engaging with more closed-mind people, users are more likely to engage with disagreeable information. And these users typically passively encountered, rather than actively sought, disagreeable information. This emphasizes the significance of recommender systems that distribute the information to users. 
Also, \citet{flaxman2016filter} conduct extensive studies on online news consumption to investigate filter bubbles. They found that the majority of consumption behavior mimicked traditional reading habits, i.e., people liked visiting the homepage and receiving information passively.

In this paper, we focus on the filter bubbles in the context of recommender systems that determine the content and information to expose to users. 
In a recommender, when a user is stuck in a filter bubble, the algorithm will show the user limited information that conforms with their preexisting belief or interest.

In recommendation, many works systematically analyzed the causes of filter bubbles. There are in general two causes. The first cause may be users' personal intention \cite{how_youtube,Auditing_youtube,Measuring_misinfo,WWW21fb}. 
For example, \citet{WWW21fb} conducted simulation studies on news recommendations and found that users with more extreme preferences are more liable to be stuck in filter bubbles. Additionally, some researchers investigate on YouTube, the most popular video-sharing platform, how and why users fall in the misinformation filter bubble, i.e., trusting inaccurate information or extreme content with controversial topics \cite{how_youtube,Auditing_youtube,Measuring_misinfo}. \citet{Measuring_misinfo} conducted large audit experiments to investigate whether personalization (i.e., age, gender, geo-location, or watch history) contributes to amplifying misinformation. Their results reveal that these personalized demographics do not significantly amplify the bubble for new-coming users. However, once these users once develop a watch history of misinformation, these demographics can exert an effect. This phenomenon brings us to the second cause of the filter bubble.

The second cause is the overexploitation behavior of the recommendation algorithm, which results in the overexposure of a certain subset of items. Recent empirical studies verified that the formation of filter bubbles is mainly due to the model emphasizing certain items with dominant user interest \cite{recsys21best,WWW21fb}. This kind of filter bubble can have a pernicious effect on user satisfaction. As demonstrated by \citet{herlocker2004evaluating}, recommending already familiar items can trigger users' unsatisfactory results, due to the fact that users like novelty and serendipity. 

It is necessary to ameliorate the filter bubble in recommender systems. Existing methods that aim to combat filter bubbles focus on improving users' awareness of diverse social opinions \cite{fb_awreness,echo_1}, enhancing models' diversity \cite{liu2020diversified,Deconstructing_fb,echo_2}, serendipity \cite{combating_fb,neural_serendipity}, fairness \cite{AAAI2020burstingfb} of recommendation results, correcting model behavior by watching debunking content \cite{recsys21best}. However, these strategies focus on the static recommendation setting, where the dynamic nature of filter bubbles is hard to model. In contrast, we capture the overexposure effect, i.e., the main symptom or proxy of filter bubbles, and solve the problem in interactive recommendation.

% \subsection{Offline Reinforcement Learning}
% Reinforcement learning learns a policy that can make dynamic decisions to meet different requirements in changing scenarios, 

\subsection{Causality-enhanced Recommendation}
Recently, causal inference (CI) has drawn a lot of attention in neural language processing (NLP) \cite{CIinNLP}, computer vision (CV) \cite{lopez2017discovering,wang2020visual}, and recommender system (RS) \cite{expomf,rs_treatment,Causalembedding_recsys18,zhangyang,yang2021top,Counterfactual-Data-Augmented}. Instead of exploiting the correlation relationships between input and output by feeding data to the black-box neural networks, CI explicitly models the causal mechanism among variables \cite{pearl2009causality,rubin2020book}. 
In recommender systems, CI can be a powerful tool for addressing various biases in the data or the learning process, e.g., selection bias and popularity bias \cite{chen2020bias}.
The most naive way to estimate the missing values is the direct method (DM) or the error-imputation-based (EIB) estimator \cite{wang2019doubly}, the idea is to use a hyper-parameter $\gamma$ to impute all missing values. So we can learn our method for these positions using $\gamma$ instead of zero. This is a very coarse-grained solution. 
Alternatively, many studies tried to estimate the unbiased user preference based on inverse propensity scoring (IPS) \cite{saito_unbiased,rs_treatment}, which is an effective CI-based method. Intuitively, when a user has a lower probability of seeing an item, we should assign increased importance to this sample and vice versa. However, IPS-based causal methods suffer from the high variance issue and it is difficult to estimate the propensity score \cite{saito_unbiased}. 
Therefore, there are methods that combine the EIB and IPS and propose the doubly robust estimator and then show the effectiveness \cite{wang2019doubly}. However, these methods have a common problem: it is hard to obtain the precise estimator, i.e., the hyper-parameter $\gamma$ in EIB and the propensity score in IPS.

Recently, researchers have followed Pearl's causal inference framework \cite{pearl2009causality} in RS \cite{zhangyang,yang2021top,xu2021causal,wenjiekdd21}. Generally, they organize the relationship of the variables as a triangle structure: a cause, an effect, and a confounder \cite{zhangyang,10.1145/3240323.3240370,10.1145/3383313.3412261}. Because of the existence of the confounder, there is a spurious relationship between the cause and the effect. Therefore, we need to cut off the path and remove the effect from the confounder in the inference stage \cite{pearl2009causality}. For example, \citet{10.1145/3383313.3412261} treat the features of users and items as the confounder and derive the unbiased inference by re-weighting the samples according to the features.

Another branch of work formulates the inference problem as a counterfactual learning framework, where the counterfactual variable refers to the potential outcome of the unobserved data \cite{clicks_can,wang2020information}. By explicitly computing the potential outcome, we can conclude whether a treatment is effective, i.e., whether taking the treatment makes a huge difference compared with doing nothing.

In this work, we generally summarize the procedure of most of the work as follows:
1) Constructing a causal graph, i.e., a representation of the structural causal model (SCM) that describes the causal relationship among the related variables.
2) In the learning stage, fitting an unbiased model (e.g., implemented as a neural network) on the training dataset based on the proposed causal graph.
3) In the inference stage, actively changing certain variables (called \emph{intervention}) according to the certain requirement, then predicting the unbiased result of the target variable.
In this work, we use this framework to explicitly model the overexposure effect of the recommender system in the filter bubble problem.

% because it enables the model to remove the effect of confounding factors \cite{wenjiekdd21,xu2021causal}, or answer the counterfactual question \cite{zhangyang} in recommendation.

% Hence, the ``What if''-based counterfactual question can be naturally answered by the intervention techniques \cite{pearl2009causality}. 
% For example, \citet{yang2021top} used the CI-based learning to answer the counterfactual question: ``what would be the user’s decision if the recommended items had been different?'' They learned a robust recommender by training on the generated hard samples based on the structural equation models (SEMs). 
% Different from almost all existing studies which focus on using CI to solve the bias problems in the static recommendation \citet{chen2020bias}. We first conduct the CI learning on interactive recommendation to help the policy better capture users' dynamic satisfaction.

\subsection{Offline RL for Recommendation}
% \subsection{Development of Interactive Recommendations}
% Most studies in recommendation pay attention to the static recommender model that is developed on a collected user feedback . 
% % Benefiting from colossal amount of collected data, researchers can easily learn a static user preference model. 
% When served the static model online, the model parameters will be frequently updated once receiving user feedback.

% In the online interaction between users and a recommender system, we often use sequential decision making algorithms to increase sales or enhance user satisfaction.

% Interactive recommender systems further introduce a reinforcement learning policy that can make decisions in the interaction automatically to pursue the desired long-term target \citep{xiaocong2021survey,afsar2021reinforcement}. However, learning a policy online is impractical as it is too slow and will hurt user experience. The intuitive solution is to leverage the offline reinforcement learning \cite{levine2020offline} to make use of the historical data. Commonly used offline RL strategies include the model-based RL methods and off-policy evaluation (OPE) learning \citep{swaminathan2015counterfactual}. The former one has the high-variance problem and the latter suffers from bias issues. Actually, bias issues are also important in recommendation. Researchers have been detecting and combating various biases in recommendation for a long time \cite{chen2020bias}. Here, we briefly summarized in \mytable{summary} six types of recommenders with respect to three dimensions: 

Recommender systems are designed to enhance user satisfaction or increase sales when serving online users. We usually consider recommendation as a sequential decision-making process, where the recommender policy decides to make recommendations according to previous user feedback. Common studies on static recommendation models (e.g., DeepFM \cite{DeepFM} and LightGCN \cite{lightgcn}) only pay attention to improving the performance in a single-round recommendation. However, this solution often assumes that the recommendation follows the I.I.D. assumption, i.e., each item or recommendation is independent and identically distributed, which is not true in many cases. 
For example, in slate recommendation or bundle recommendation, the conversion rate of an item does not solely depend on itself. If it is surrounded by similar but expensive items, the conversion rate can increase. This phenomenon is known as the decoy effect \cite{decoyRS}. Besides, many items have future impacts rather than instantaneous rewards, e.g., recommending a high-priced item may not result in an instant consumption behavior now, but it can leave the user an impression that there are high-quality items in this platform and this impression can result in transactions in the future when users have the ability to consume the expensive items. The overexposure effect of filter bubbles can also have an effect on users' experience, but the effect is pernicious, i.e., it can negatively affect users' willingness to continue to use and trust this recommender system. 

To directly model the long-term success in the multi-round decision-making problem, we can adopt the interactive recommender system (IRS) based on reinforcement learning (RL) \cite{gao2023alleviating,wang2022best,liu2018deep,sigir20_interactive,zhang2019reward,NICF,lixin_kdd,liu2020diversified,gao2021advances,gao2021tutorial,cai2023reinforcing,liu2023exploration,ResAct,cai2023two}.
The object of reinforcement learning is to maximize the cumulative reward $J\left(\pi_{\theta}\right)$ as:
\begin{equation}
    J(\pi)=\mathbb{E}_{\tau \sim p_{\pi}(\tau)}\left[\sum_{t=0}^{H} \gamma^{t} r\left(\mathbf{s}_{t}, \mathbf{a}_{t}\right)\right]. 
\end{equation}
where $r\left(\mathbf{s}_{t}, \mathbf{a}_{t}\right)$ is a reward given from the environment when the agent makes action $\mathbf{a}_{t}$ at state $\mathbf{s}_{t}$, $\gamma\in(0,1]$ is a scalar discount factor. The trajectory $\tau$ is a sequence of states and actions of length $H$, given by $\tau=\left(\mathbf{s}_{0}, \mathbf{a}_{0}, \ldots, \mathbf{s}_{H}, \mathbf{a}_{H}\right)$. $p_\pi(\tau)$ is a trajectory distribution for a given Markov decision process (MDP) and the policy $\pi$ is written as follows:
\begin{equation}
    p_{\pi}(\tau)=d_{0}\left(\mathbf{s}_{0}\right) \prod_{t=0}^{H} \pi\left(\mathbf{a}_{t} \mid \mathbf{s}_{t}\right) T\left(\mathbf{s}_{t+1} \mid \mathbf{s}_{t}, \mathbf{a}_{t}\right).
\end{equation}
where $T\left(\mathbf{s}_{t+1} \mid \mathbf{s}_{t}, \mathbf{a}_{t}\right)$ is a state transition probability that describes the dynamics of the environment system.

Though effective, learning an IRS policy with online users is impractical as it is too slow and will hurt user experience \cite{gilotte2018offline}. On the other hand, the recorded recommender logs and offline user feedback are 
easier to obtain \cite{kiyohara2021accelerating}. Therefore, it is natural to think of learning a policy on the offline data, which is the core idea of offline RL \cite{levine2020offline}.

Commonly used offline RL strategies in recommendation include off-policy evaluation (OPE) learning \citep{swaminathan2015counterfactual,minmin_topK} and model-based RL methods \cite{Pseudo-Dyna-Q,chen2019generative}. OPE-based methods estimate the cumulative reward of the target policy $\pi_\theta$ using the weighted data from logged policy $\pi_\beta$. A classical importance sampling estimator \cite{precup2000eligibility} is listed as follows:
\begin{equation}
    \begin{aligned}
    J\left(\pi_{\theta}\right) &=\mathbb{E}_{\tau \sim \pi_{\beta}(\tau)}\left[\frac{\pi_{\theta}(\tau)}{\pi_{\beta}(\tau)} \sum_{t=0}^{H} \gamma^{t} r(\mathbf{s}, \mathbf{a})\right] \\
    &=\mathbb{E}_{\tau \sim \pi_{\beta}(\tau)}\left[\left(\prod_{t=0}^{H} \frac{\pi_{\theta}\left(\mathbf{a}_{t} \mid \mathbf{s}_{t}\right)}{\pi_{\beta}\left(\mathbf{a}_{t} \mid \mathbf{s}_{t}\right)}\right) \sum_{t=0}^{H} \gamma^{t} r(\mathbf{s}, \mathbf{a})\right] \approx \sum_{i=1}^{n} w_{H}^{i} \sum_{t=0}^{H} \gamma^{t} r_{t}^{i}
    \end{aligned}
\end{equation}
where $w_{t}^{i}=\frac{1}{n} \prod_{t^{\prime}=0}^{t} \frac{\pi_{\theta}\left(\mathbf{a}_{t^{\prime}}^{i} \mid \mathbf{s}_{t^{\prime}}^{i}\right)}{\pi_{\beta}\left(\mathbf{a}_{t^{\prime}}^{i} \mid \mathbf{s}_{t^{\prime}}^{i}\right)}$ and $\left\{\left(\mathbf{s}_{0}^{i}, \mathbf{a}_{0}^{i}, r_{0}^{i}, \mathbf{s}_{1}^{i}, \ldots\right)\right\}_{i=1}^{n}$ are $n$ trajectory samples from the historical policy $\pi_\beta$. However, the estimator in OPE-based methods often suffers from the high variance problem due to the divergence between the two distributions in $w_{t}^{i}$. 

The model-based methods attempt to estimate the transition function $T(\mathbf{s}_{t+1}|\mathbf{s}_{t},\mathbf{a}_{t})$ and the reward function $r_{t}$ of the environment. From the estimated model, the RL agent can plan its trajectory accordingly on the predicted states and rewards, rather than learning rigorously on the historical trajectories. The model-based method can avoid the high-variance problem in OPE-based methods, however, it suffers from the bias issue in estimating the model. In recommender systems, various biases were specified and investigated \cite{chen2020bias}. In this work, we specify the bias as to whether the model considers the user feeling bored in the interaction.

 % one has the high-variance problem and the latter suffers from bias issues \cite{levine2020offline}. 
% Actually, bias issues are also important in recommendation. Researchers have been detecting and combating various biases in recommendation for a long time \cite{chen2020bias}.

As mentioned above, CI is a powerful tool to address the bias problem in recommender systems. Hence, we combine the model-based offline RL and the CI technique to develop an unbiased IRS that can recognize and address the filter bubble problem on the offline data. We choose to use the model-based offline RL because it has strong advantage of being sample efficient \cite{nagabandi2018neural,deisenroth2013gaussian}, which is crucial in recommender system where the data is highly sparse and expensive to collect. Besides, we can directly model the bias in the extracted transition probability and reward function via a causality-enhanced model, which is also a main contribution of this work.

Here, we briefly summarized in \mytable{summary} six types of recommenders with respect to three dimensions: 
(1) whether the system explicitly builds a user model trying to capture real user preference, 
(2) whether the system considers debiasing, and 
(3) whether the system has an RL-based policy. 
% Note that our proposed CIRS model is an unbiased model-based IRS. It is unbiased since it can recognize the effect of item overexposure on user satisfaction. 
Note that there are two other works in the same category as our CIRS, but they are not designed for the filter bubble problem. 
In addition, the unbiasedness of them is not referred to the ability to address a certain bias effect in the recommendation problem, but to make the estimation more accurate.

\begin{table}[t]
\caption{Six Types of Recommender Systems}
\label{tab:summary}
% \vspace{-2mm}
\tabcolsep=8pt
% \scriptsize
% \tabcolsep=2pt
% \footnotesize
\renewcommand\arraystretch{1.1}
\begin{tabular}{@{}lcccc@{}}
\toprule
                   & \textbf{User Model}                         & \textbf{Debiasing}            & \textbf{RL-based}                           & \textbf{Publications}         \\ \toprule
Static RS          & \green{\small{\Checkmark}}                         &                      &                                    &\citep{MFmodel,DeepFM,lightgcn,yu2019generating}                      \\ \midrule 
Unbiased static RS & \green{\small{\Checkmark}}                         & \green{\small{\Checkmark}}           &                                    &\citep{saito_unbiased,rs_treatment,KDD21popularitybias,li2021causal,wenjiekdd21,zhangyang,DICE,bias_general}                   \\ \midrule
Traditional IRS    &                                    &                      & \green{\small{\Checkmark}}                         & \citep{liu2018deep,Xinxin,sigir20_interactive,zhang2019reward,NICF,lixin_kdd,embedding_IRS,liu2020diversified} \\ \midrule
Model-based IRS    & \green{\small{\Checkmark}}                         &                      & \green{\small{\Checkmark}}                         &   \citep{Pseudo-Dyna-Q,xiangyu_model_based,xiangyuwww21,nips_model_bai,chen2019generative,wang2021AAAI}                  \\ \midrule
OPE-based IRS      &                                    & \green{\small{\Checkmark}}           & \green{\small{\Checkmark}}                         &   \citep{swaminathan2015counterfactual,minmin_topK,ie2019youtube,off_two_stage,xiao2021general,jagerman-when-2019,KDD20_counterfactual,Pessimistic2021recsys}\\ \midrule
\begin{tabular}[c]{@{}c@{}}Unbiased model-\\ based IRS\end{tabular} & \green{\small{\Checkmark}}                   & \green{\small{\Checkmark}}           & \green{\small{\Checkmark}}                   &  \begin{tabular}[c]{@{}c@{}}\citep{Keeping-recsys,minmin_user_model,gao2023alleviating}\\\textbf{CIRS (Ours)}\end{tabular} \\ \bottomrule
\end{tabular}
% \vspace{-3mm}
\end{table}
\section{Prerequisites}
\label{sec:pre}
In this section, we introduce the problem definition and the empirical analyses of the real-world data from Kuaishou.

\subsection{Problem Definition}

We denote the user set as $\mathcal{U}$ and item set as $\mathcal{I}$. 
The set of all interaction sequences of a user $u\in\mathcal{U}$ can be denoted as $\mathcal{D}_u = \{\mathcal{S}_u^1,\mathcal{S}_u^2,\cdots,\mathcal{S}_u^{\left|\mathcal{D}_u\right|}\}$. Each $\mathcal{S}_u^k\in\mathcal{D}_u$ is the $k$-th interaction sequence (i.e., trajectory) recording a complete interaction process: $\mathcal{S}_u^k=\{(u, i_l, t_l)\}_{1\leq l \leq \left|\mathcal{S}_u^k\right|}$, where the user $u$ begins to interact with the system at time $t_1$ and quits at time $t_{|\mathcal{S}_u^k|}$, and $i_l \in \mathcal{I}$ is the recommended item at time $t_l$. Let $\mathbf{e}_u \in \mathbb{R}^{d_u}$ and $\mathbf{e}_i \in \mathbb{R}^{d_i}$ be the feature representation vectors of user $u$ and item $i$, respectively.
For the system, the task is to recommend items to users based on their preferences and interaction history. This process can be cast as a reinforcement learning problem, whose key components are summarized as follows:
\begin{itemize}
\item \textbf{Environment}. The agent works in the environment where the states and rewards are generated. Here, the environment is the user (either an online real user or a simulated one) who can choose to rate the recommended items from the system or quit the interaction.
\item \textbf{State}. The system maintains a state $\mathbf{s}_t \in \mathbb{R}^{d_s}$ at time $t$ is regarded as a vector representing information of all historical interactions between user $u$ and the system prior to $t$. In this paper, we obtain the $\mathbf{s}_t$ by using the Transformer model \cite{transformer}.
\item \textbf{Action}. The system makes an action $a_t$ at time $t$ is to recommend items to user $u$. Let $\mathbf{e}_{a_t} \in \mathbb{R}^{d_a}$ denote the representation vector of action $a_t$. In this paper, each action $a_t$ recommends only one item $i$. Hence, we have $\mathbf{e}_{a_t} = \mathbf{e}_i$.
\item \textbf{Reward}. The user $u$ returns feedback as a reward score $r_t$ reflecting its satisfaction after receiving a recommended item $i$. 
The reward can also be the \emph{counterfactual satisfaction} predicted by the causal user model $\phi_M$ instead of real users' feedback.
\item \textbf{State Transition}. After the agent makes an action $a_t$ and the user gives a reward $r_t$, the state $\mathbf{s}_t$ will be updated to $\mathbf{s}_{t+1}$ according to a state transition probability $T(\mathbf{s}_{t+1}|\mathbf{s}_t, a_t)$. In our work, this transition probability is modeled by a state tracker implemented on the Transformer model \cite{transformer}. 
% In our method, the reward in the RL planning stage is given by the causal user model $\phi_M$. Since the reward is given by $\phi_M$ instead of the user's feedback, we name it as \emph{counterfactual satisfaction}.
\item \textbf{Policy}. The key task of the system is to optimize a target policy $\pi_\theta = \pi_\theta(a_t|\mathbf{s}_t)$ that represents the probability of making an action $a_t$ conditioned on the state $\mathbf{s}_t$. It decides how to generate an item $i$ to recommend and is usually implemented as a fully-connected neural network.
\end{itemize}

\medskip
To make full use of the historical data, we base our method CIRS on the offline RL framework. Learning CIRS requires three recurrent steps (as illustrated by three color blocks in \myfig{method}):
\begin{enumerate}
\item Training a causal user model $\phi_M$ on historical interaction data $\{(u, i, r)\}$ to estimate not only user interest but also the effect of item overexposure.
\item Using the learned causal user model $\phi_M$ (instead of real users) to train policy $\pi_\theta$. In each interaction loop, $\phi_M$ samples a user $u$ to interact with $\pi_\theta$. When $\pi_\theta$ makes an action $a_t$ (recommends $i$), $\phi_M$ provides a \emph{counterfactual satisfaction} as the reward $r$. Intuitively, if $\pi_\theta$ has made similar recommendations before, $\phi_M$ shrinks the reward $r$.
\item Serving the learned policy $\pi_\theta$ to real users and evaluating the results in the interactive environment. When the interaction ends, we save the log $\{u, i, r, t\}$ to historical data for the purpose of future learning.
\end{enumerate}
The three steps can be conducted repeatedly to continuously improve $\phi_M$ and $\pi_\theta$.

\subsection{Field Study of Overexposure Effect on User Satisfaction}
\label{sec:verification}

As mentioned above, besides users' responsibility, the cause of filter bubbles is that the model incessantly recommends similar items to users. We suppose this is pernicious as users may not feel comfortable under such circumstances. Thus, we present a hypothesis as follows:

\begin{hyp}
\label{hyp:hyp}
Users' satisfaction will drop if the recommended items (or similar items) are repeatedly exposed and recommended to them in a short time.
\end{hyp}

To verify this hypothesis, we conduct empirical studies on real data from Kuaishou, a video-sharing mobile App. We chose $7,176$ users from the video-watching user pool of the platform. We filter and collect their interaction history from August 5th, 2020 to August 11th, 2020. There are $34,215,294$ views in total and $3,110,886$ videos were watched. Each item is tagged with at least one and at most four-category tags, and there are $31$ category tags in total.

During watching a video, users can choose to quit watching at any time by scrolling down to the next video or leaving the video-playing interface. Each video has a comments section that users can enter by clicking the ``comment'' button. If users are interested in a video, they will stay watching it for a longer time or enter its comments section. Therefore, we design two key metric indicators to reflect user satisfaction: One is the time duration staying in the comments section, and the other is the video watching ratio, which is the ratio of viewing time to the total video length.
% Both of the two indicators are directly proportional to user satisfaction.

\begin{figure}[!t]
\centering
\includegraphics[width=.7\linewidth]{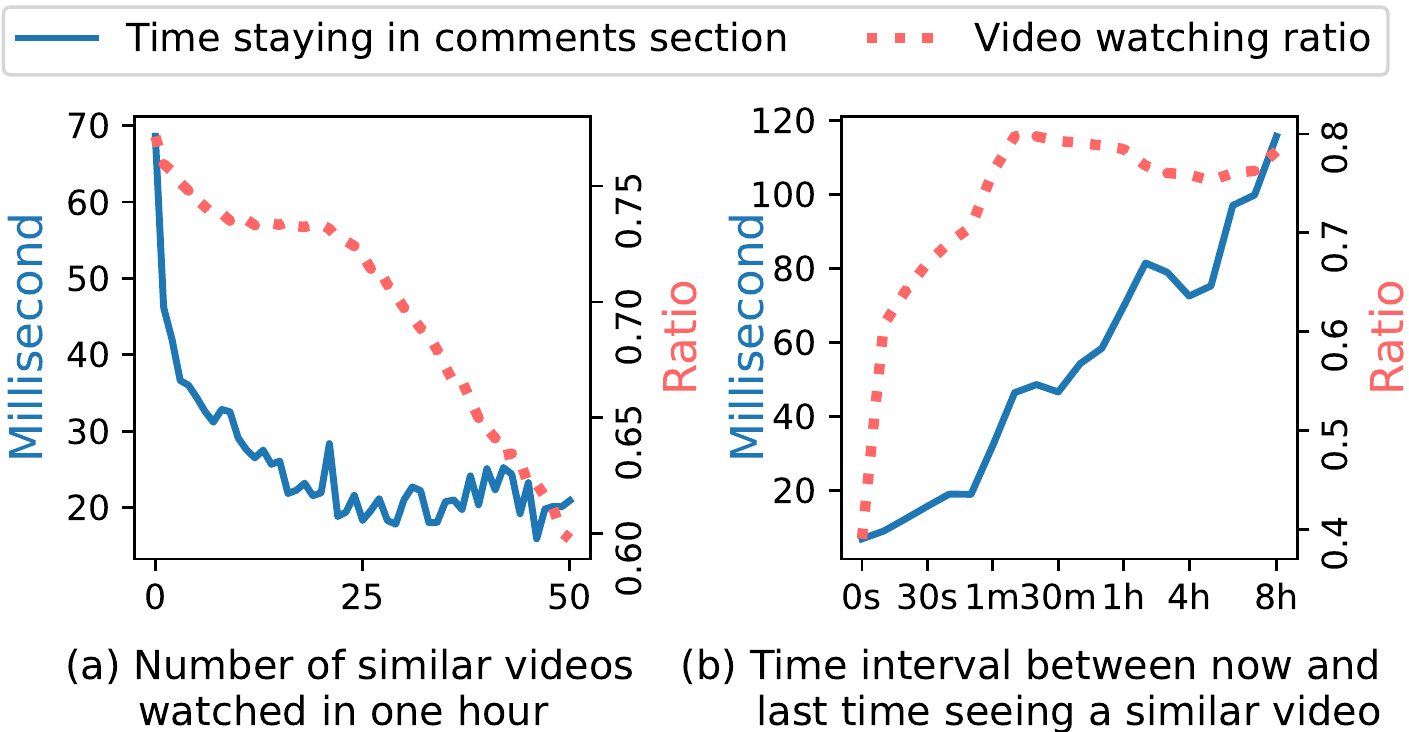}
% \vspace{-2mm}
\caption {Empirical studies of exposure effect on Kuaishou. Statistics of two metrics vary with the exposure effect.}
% \vspace{-2mm}
\label{fig:statistics}
\end{figure}

To show how item exposure affects user satisfaction, we study how the two indicators change with the degree of overexposure effect. Specifically, we group all collected views with respect to (a) the number of videos with the same tag watched in one hour, or (b) the time interval between now (watching this video) and the last time viewing a video with the same tag. Then we compute the average values of the mentioned indicators in every group. The results are shown in \myfig{statistics}. Two observations are found:
\begin{obs}
User satisfaction towards a recommended item drops when the system increases the number of similar items in recent recommendations.
\end{obs}
\begin{obs}
User satisfaction toward a recommended item drops as the time interval between two similar items is shortened.
\end{obs}

The result is statistically significant since even the smallest group contains enough points: $31,277$ points for the group at $x=50$ in \myfig{statistics} (a) and $76,303$ points for the group at $x=(7h\sim8h)$ in \myfig{statistics} (b).

In our previous work \cite{gao2023alleviating}, an empirical analysis also demonstrated a similar phenomenon: user satisfaction decreases as the repetition rate of items or categories increases.
Therefore, \myhyp{hyp} is empirically proved. This suggests that the filter bubble, i.e., the overexposure effect of the recommendation algorithm, does have a pernicious impact on user satisfaction. Consequently, we can design a ``feel bored then quit'' exit mechanism in the constructed environments as a reflection of user behavior, which enables us to simulate the effect of filter bubbles effectively. We will illustrate it in \mysec{exp}.
\section{Proposed Methods}
\label{sec:method}

In this section, based on the observations obtained in the field studies, we propose a counterfactual interactive recommender system (CIRS) that leverages causal inference in offline RL. 
We first introduce CIRS's three main modules in the offline RL framework, then we describe how to leverage causal inference to disentangle the causal effects on user satisfaction.

\subsection{Offline RL-based Framework}
\begin{figure}[!t]
\centering
\includegraphics[width=.6\linewidth]{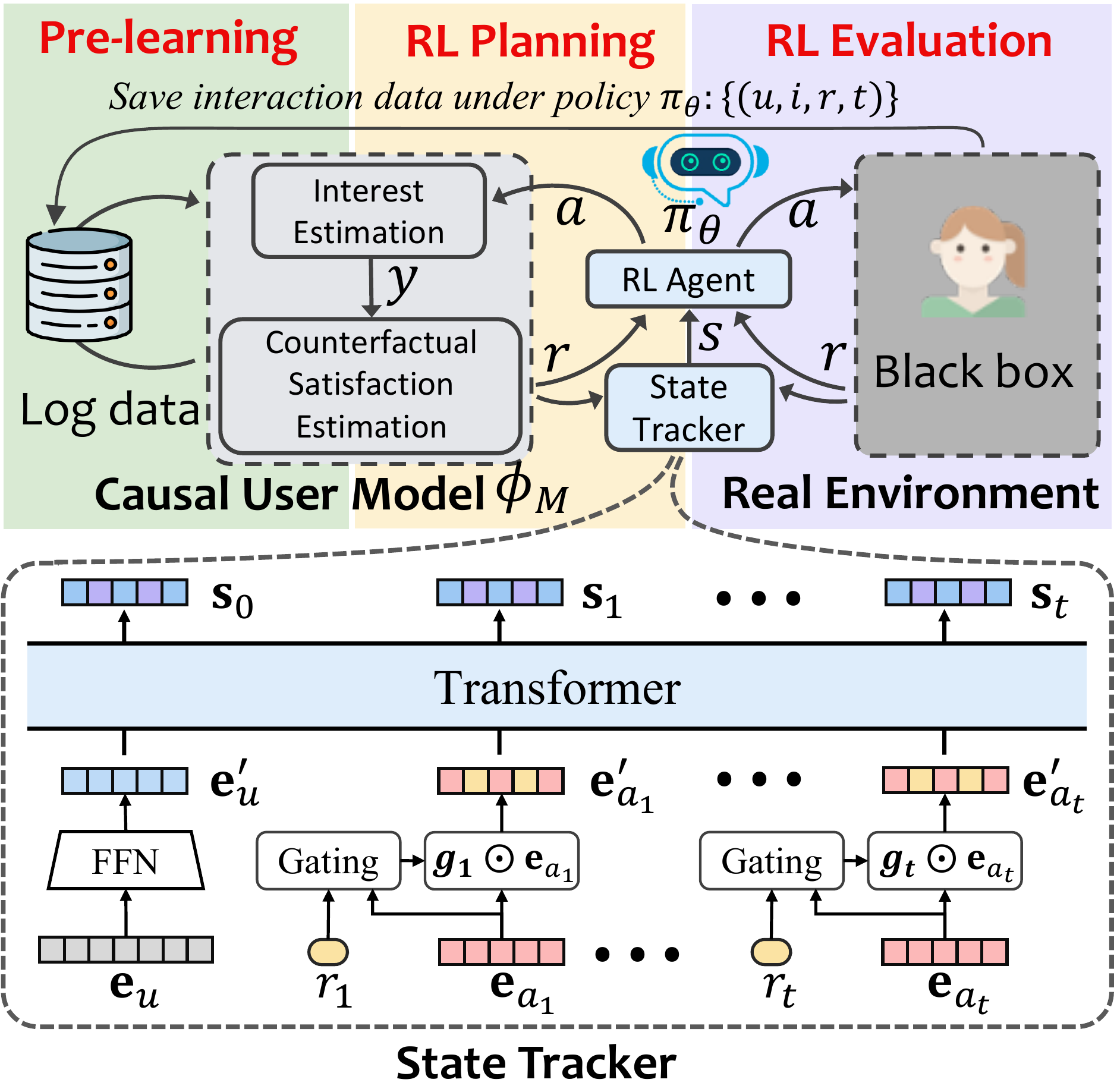}
% \vspace{-2mm}
\caption {Learning Framework of Counterfactual IRS.}
% \vspace{-2mm}
\label{fig:method}
\end{figure}

We base our CIRS model on offline RL, where we can utilize a large amount of offline data to train the interactive recommender system. The framework of CIRS is illustrated in \myfig{method}. It contains three stages (shown in three color blocks): pre-learning stage, RL planning stage, and the RL evaluation stage. The functions of these three stages correspond to the three: 
(1) pre-learning the user model $\phi_M$ via supervised learning, 
(2) using the learned user model $\phi_M$ to learn RL policy $\pi_\theta$ by providing \emph{counterfactual satisfaction} as the reward, and 
(3) evaluating policy $\pi_\theta$ in the real environment.
Especially, the real environment can either be the real-world online environment or the simulation environment that can reflect real user behavior.
% However, it is impractical to evaluate the policy online with real users because the interaction is slow and this process hurts user experiences \cite{gao2021advances,gilotte2018offline,schnabel2018short}. There
Next, we separately introduce the three main components in CIRS: causal user model $\phi_M$, the state tracker module, and RL-based interactive policy $\pi_\theta$.

\subsubsection{Causal User Model}
% The causal user model $\phi_M$ learns user interest based on the historical data. It can correctly capture the effect of item overexposure on user experience. There are two sub-modules in the user model: an interest estimation module and a casual intervention module.
The causal user model $\phi_M$ learns user interest based on the historical data and provides the counterfactual satisfaction $r$ for the RL planning stage. It aims to explicitly disentangle the causal effect by correctly modeling the effect of item overexposure on user satisfaction. There are two sub-modules in the user model: an interest estimation module designed for computing the user's intrinsic interest $y$, and a counterfactual satisfaction estimation module capturing how the overexposure effect affects user satisfaction $r$. 
The details of this component will be reported in \mysec{causal},
where we will formally introduce the causal view of the recommender.

\subsubsection{Transformer-based State Tracker}
The state $\mathbf{s}_t$ in the interactive recommendation should include all critical information at time $t$ for policy $\pi_\theta$ to make decisions. That includes: the feature vector $\mathbf{e}_u$ of the user $u$, the feature vectors of the recently recommended items, i.e., actions of the system in the interaction loop $\{\mathbf{e}_{a_1},\cdots,\mathbf{e}_{a_t}\}$, and the user's feedback towards them $\{r_1,\cdots,r_t\}$. 
In order to automatically extract key information from  these vectors, we use the Transformer model \cite{transformer} to derive $\mathbf{s}_t$ as illustrated in \myfig{method}.
Transformer is a state-of-the-art sequence-to-sequence model with an attention mechanism that can capture the dependence between current input and previous input sequences \cite{zhang2022mixhead}. In this work, we use only a two-layer encoder of Transformer. 
Since the input is generated sequentially, we need to add a mask to prevent future information leaking into each state of the sequence.

We further use a gate mechanism to filter information from the action $\mathbf{e}_{a_t}$ and user feedback $r_t$. Hence, the input for Transformer at time $t$ is $\mathbf{e}^\prime_{a_t}:=\boldsymbol{g}_{t}\odot\mathbf{e}_{a_t}$. Where `$\odot$' denotes the element-wise product, and the gating vector $\boldsymbol{g}_{t}$ is computed as:
\begin{equation}
\boldsymbol{g}_{t}=\sigma\left(\mathbf{W} \cdot \operatorname{Concat}\left(\mathbf{r}_{t}, \mathbf{e}_{a_t}\right)+\mathbf{b}\right),
\end{equation}
where $\mathbf{W} \in \mathbb{R}^{d_s \times (1+ d_a)}$ and $\mathbf{b} \in \mathbb{R}^{d_s}$ refers to the weight matrix and bias vector, respectively. $\operatorname{Concat}(\cdot)$ is the vector concatenation operator. Besides, we use a feed-forward network ($\operatorname{FFN}$) to transform the representation vector of user $u$ from $\mathbf{e}_u \in \mathbb{R}^{d_u}$ to $\mathbf{e}^\prime_u \in \mathbb{R}^{d_s}$ to let it lie in the same space 
of $\mathbf{e}^\prime_{a_t}$.

The feature vectors, the parameters in Transformer and the gate are initialized randomly and trained in the end-to-end manner using the gradients passed from the RL model.

It is noteworthy that the Transformer model and the gate mechanism can be replaced by other architectures, since there are other ways to extract and combine the information from the input sequence, such as a recurrent neural network (RNN)-based model \cite{10.1145/3437963.3441824,chen2019large,wang2022best}.
Besides, The parameters in Transformer and the gate mechanism are fixed during the RL policy updating its parameters. Afterward, they are updated by another Adam optimizer \cite{adam}, using the gradient that was back-propagated from the RL policy.

\subsubsection{RL-based Interactive Recommendation Policy}
We implement our interactive recommendation policy $\pi_\theta$ as the PPO algorithm \cite{ppo}. PPO is a powerful on-policy reinforcement learning algorithm based on actor-critic framework \cite{a2c} and the trust region policy optimization (TRPO) algorithm \cite{TRPO}.
It can work in both discrete and continuous state space and action space.
We aim to maximize the cumulative user satisfaction of the long-term interaction, which can be realized by maximizing the objective function of PPO:
\begin{equation}
\label{eq:L_rl}
\mathbb{E}_{t}\left[\min \left(\frac{\pi_{\theta}\left(a_{t} \mid \mathbf{s}_{t}\right)}{\pi_{\theta_{\text{old}}}\left(a_{t} \mid \mathbf{s}_{t}\right)} \hat{A}_t, \operatorname{clip}\left(\frac{\pi_{\theta}\left(a_{t} \mid \mathbf{s}_{t}\right)}{\pi_{\theta_{\text{old}}}\left(a_{t} \mid \mathbf{s}_{t}\right)}, 1-\epsilon, 1+\epsilon\right) \hat{A}_t\right)\right],
\end{equation}
where $\epsilon$ is the hyperparameter that controls the maximum percentage of change that can be updated at one time. The function $\operatorname{clip}(x,a,b)$ clips the variable $x$ in the range of $[a,b]$. $\theta_{\text{old}}$ is the policy before updating, i.e., the interaction data are generated under policy $\theta_{\text{old}}$. And the advantage function $\hat{A}_t$ is implemented as the generalized advantage estimator (GAE) \cite{gae}
given as follows:
% which can be generally understood as a quantity that is proportional the normalized cumulative rewards of a sequence $\sum_{l=0}^{|\mathcal{S}|} \gamma^{l} r_{t+l}$. For more details, we refer the audiences to the original PPO algorithm \cite{ppo}.
\begin{equation}
    \hat{A}_t := \hat{A}_{t}^{\mathrm{GAE}(\gamma, \lambda)} :=\sum_{l=0}^{\infty}(\gamma \lambda)^{l} \delta_{t+l}^{V}
\end{equation}
where $\lambda \in [0,1]$ is a hyperparameter making a compromise between bias and variance. $\delta_{t}^{V}$ is defined as $\delta_{t}^{V}=r_{t}+\gamma V\left(s_{t+1}\right)-V\left(s_{t}\right)$, i.e., the TD residual of the approximate value function $V$ with discount $\gamma$ \cite{sutton2018reinforcement_book}. The value function $V$ is defined as:
\begin{equation}
    V\left(s_{t}\right):=V^{\pi_\theta, \gamma}\left(\mathbf{s}_{t}\right)
    :=\mathbb{E}_{\mathbf{s}_{t+1: \infty}, a_{t: \infty}}\left[\sum_{l=0}^{\infty} \gamma^{l} r_{t+l}\right]
\end{equation}
It should be noted that the reward term $r_t$ in $\hat{A}_t$ is the \emph{counterfactual satisfaction} given by the causal user model $\phi_M$ instead of immediate users' feedback. 

At last, to learn a good policy, we need the \emph{counterfactual satisfaction} given by $\phi_M$ to be as correct and constructive as possible. Next, we will introduce how to build a causal user model $\phi_M$ to capture the overexposure effect thus avoiding the filter bubble.

\begin{figure}[!t]
\centering
\includegraphics[width=.7\linewidth]{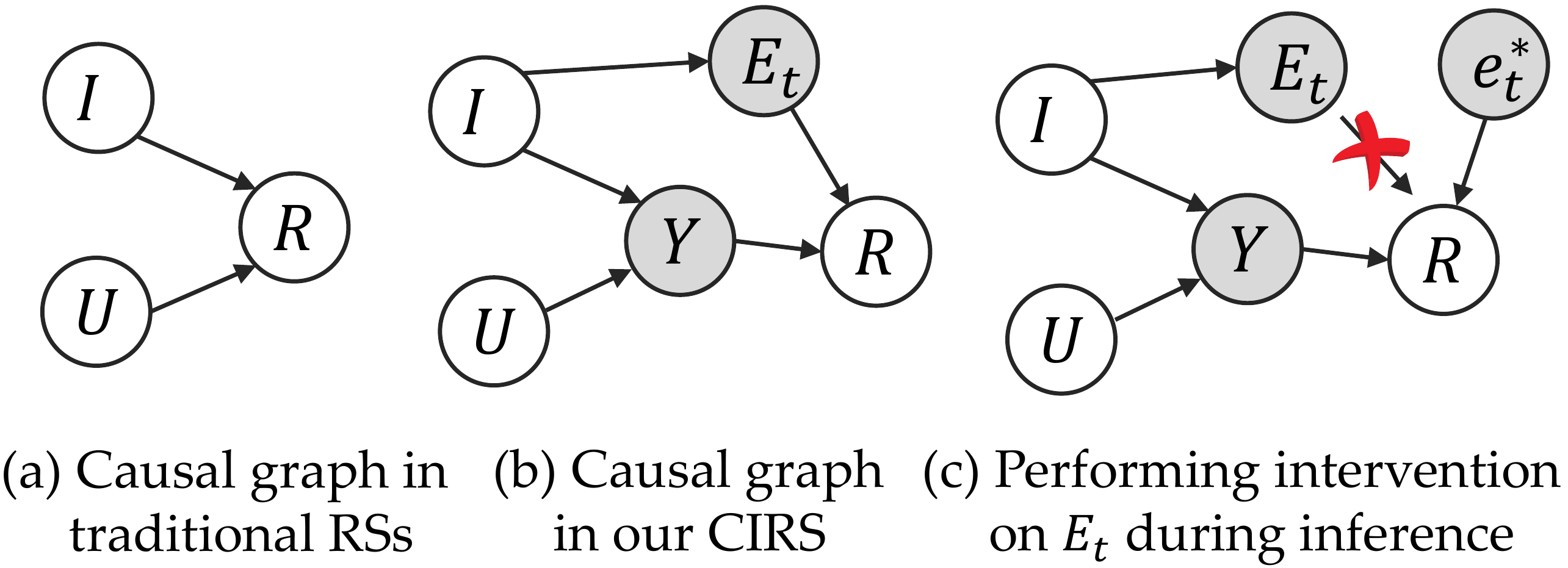}
\caption {Causal graphs of traditional IRS models (a) and the CIRS model (b-c). $U$: a user, $I$: an item, $R$: user satisfaction (quantified by feedback such as clicks or viewing time), $Y$: intrinsic interest, $E_t$ and $e^*_t$: the overexposure effect. Latent variables are shaded.}
% \vspace{-2mm}
\label{fig:causal}
\end{figure}

% \subsection{Causal View of Interactive Recommendations}
\subsection{Causal Inference-based User Satisfaction Disentanglement}
\label{sec:causal}

% \subsubsection{Causal View of User Experience Modeling}

We illustrate the causal graphs of the traditional recommender systems and our CIRS model in \myfig{causal}. 

\begin{itemize}
\item Node $U$ represents a certain user $u$, e.g., an ID or the profile feature that can represent the user.
\item Node $I$ represents an item $i$ that is recommended to user $u$.
\item Node $R$ represents user $u$'s real-time satisfaction for the recommended item $i$. It is the feedback such as a click or the video watching ratio.
% The  $r$ is used as the \emph{counterfactual satisfaction} in the offline RL-based framework.
\item Node $Y$ represents the user's intrinsic interest that is static regardless of item overexposure.
\item Node $E_t$ and $e^*_t$ represent the overexposure effect of item $i$ on user $u$. $E_t$ is a random variable and $e^*_t$ is the value of $E_t$ computed in the inference stage (i.e., the RL planning stage).
\end{itemize}

Traditional recommender systems will fit user satisfaction $R$ based on solely the feature information of user $U$ and item $I$ (\myfig{causal} (a)). This assumes that user satisfaction equals to intrinsic interest, which is improper as stated before.

Our proposed CIRS model innovatively takes into account the overexposure effect $E_t$, and disentangles the causal effect on user satisfaction $R$. Concretely, $R$ is generated from two causal paths:
\begin{enumerate}
    \item $(U, I) \rightarrow Y \rightarrow R$: This path projects user $u$ and item $i$ to their corresponding intrinsic interest ${y}_{ui}$. Then user satisfaction ${r}_{ui}$ is proportional to ${y}_{ui}$.
    \item $I \rightarrow E_t \rightarrow R$: This path captures the real-time overexposure effect $e_t(u,i)$ of an item $i$ on user $u$'s satisfaction $r$. This effect negatively affects user satisfaction.
\end{enumerate}

\smallskip\noindent\textbf{Intrinsic Interest Estimation.}
We estimate the user's intrinsic interest ${y}_{ui}$ as $\hat{y}_{ui} = f_\theta (u, i)$. The estimation model $f_\theta (u, i)$ can be implemented by almost any established recommendation model, such as DeepFM \cite{DeepFM} used in this work. We illustrated this part in the interest estimation module in \myfig{method}. 

\smallskip\noindent\textbf{Overexposure Effect Definition}
Considering that overexposure effect negatively affects user satisfaction, 
we define the overexposure effect $e_t$ of recommending an item $i$ to user $u$ at time $t$ as:
\begin{equation}
\label{eq:ee}
e_t:= e_t(u,i):= \alpha_u \beta_i \hspace{-10pt}\sum_{\left(u, i_{l}, t_{l}\right) \in \mathcal{S}_{u}^{k},t_{l}<t} \hspace{-10pt} \exp\left(-\frac{t-t_{l}}{\tau}\times \operatorname{dist}(i,i_l)\right),
\end{equation}
where $\operatorname{dist}(i,i_l)$ is distance between two items $i$ and $i_l$. $\alpha_u$ represents the \textit{sensitivity} of user $u$ to the overexposure effect, e.g., a user with a large $\alpha_u$ is more likely to feel bored when overexposed to similar content. Likewise, $\beta_i$ represents the \textit{unendurableness} of item $i$. For example, a classical music might be more endurable than a pop song, so $\beta_i$ of the classical music is smaller. $\tau$ is a temperature hyperparameter. 
We will show the relationship between the learned $\alpha_u$ ($\beta$) and the activity of a user (the popularity of an item) in \mysec{key_p}.
Intuitively, when the recommended item $i$ is close to the previously consumed items (e.g., $i_l$) of this user, i.e., $\operatorname{dist}(i,i_l)$ is small, and its recommended time $t$ is close to the recommended time of the previous ones (e.g., $t_l$ of item $i_l$), i.e., the term $(t_l - t)$ is small, then the overexposure effect $e_t(u,i)$ will be large. This means that the recommender system is introducing a filter bubble to the user.

\smallskip\noindent\textbf{User Satisfaction Estimation.}
Generally, a similar item $i_l$ (i.e., with smaller $\operatorname{dist}(i,i_l)$) that was recommended recently (i.e., with smaller $t-t_l$) contributes a larger overexposure effect to item $i$. And $e_t$ is the sum of the effect of all items recommended to the user $u$ before time $t$.
After obtaining the overexposure effect $e_t$ via \myeq{ee} and intrinsic interest $\hat{y}_{u i}$ via DeepFM, we estimate user $u$'s satisfaction on item $i$ as: 
% $\hat{y}_{u i}^{t}=\frac{\hat{r}_{u i}}{1+e_t(u,i)}.$
\begin{equation}
\hat{r}^{t}:=\hat{r}_{u i}^{t}=\frac{\hat{y}_{u i}}{1+e_t(u,i)}.
\end{equation}
Therefore, a large exposure effect $e_t(u,i)$ will diminish user satisfaction $\hat{r}_{u i}^{t}$ even with unchanged intrinsic interest $y_{ui}$. 
% This part is illustrated in the counterfactual satisfaction estimation module in \myfig{method}.
In the training stage of causal user model $\phi_M$, we minimize the objective function in the recommendation model. In experiments, we use the MSE loss for the VirtualTaobao and BPR loss for KuaiEnv:
\begin{equation}
    L_{\mathrm{MSE}}=\hspace{-10pt} \sum_{(u, i, t) \in \mathcal{D}}\hspace{-10pt}
    \left(\hat{r}_{u i}^{t}-r_{u i}^{t}\right)^2,
    \hspace{5pt}
    L_{\mathrm{BPR}}=- \hspace{-15pt} \sum_{(u, i, t) \in \mathcal{D}, j \sim p_{n}}\hspace{-15pt}
    \log \left(\sigma\left(\hat{r}_{u i}^{t}-\hat{r}_{u j}^{t}\right)\right).
\end{equation}
Where $\sigma(x)=\frac{1}{1+e^{-x}}$ is the Sigmoid function. The item $j$ is a negative instance sampled from the distribution $p_n$.

\smallskip\noindent\textbf{Counterfactual Satisfaction Estimation.}
In the RL planning stage,
when the learned causal user model $\phi_M$ interacts with the policy $\pi_\theta$, 
% the time scale of interactions and the pattern of recommendation are changed. Consequently, 
the overexposure effect $e^*_t$ now is different to $e_t$ in the pre-learning stage. Therefore, we perform the causal intervention $do(E_t=e^*_t)$ \cite{pearl2009causality} by cutting off the path $I \rightarrow E_t \rightarrow R$ as shown in \myfig{causal}(c). Unlike traditional causal methods aiming to remove the effect of confounders \cite{wang2020visual,wenjiekdd21}, we still need to model the correct overexposure effect $e^*_t$ in this stage. Note that we use the asterisk to mark out all values in this intervention stage. We compute $e^*_t$ as:
\begin{equation}
\label{eq:eestar}
e^*_t(u,i) = \gamma^*\cdot \alpha_u \beta_i \hspace{-10pt}\sum_{\left(u, i_{l}^*, t_{l}^*\right) \in \mathcal{S}_{u}^{*},t_{l}^*<t} \hspace{-10pt} \exp\left(-\frac{t-t_{l}^*}{\tau^*}\times \operatorname{dist}(i,i_l^*)\right),
\end{equation}
where $\mathcal{S}_{u}^{*}$ is the new interaction trajectory produced in the RL planning stage. $\gamma^*$ is a hyper parameter introduced to adjust the scale of the overexposure effect. We fix $\gamma$ to be $10$ throughout experiments. $\tau^*$ is the temperature hyperparmerter in the intervention stage and can have a different value with $\tau$ in \myeq{ee}.
We estimate the \emph{counterfactual satisfaction} as:
% $\hat{r}^{t*}_{u i}=\frac{\hat{y}_{u i}}{1+ e^*_t(u,i)}.$
\begin{equation}
\hat{r}^{t*}_{u i}=\frac{\hat{y}_{u i}}{1+ e^*_t(u,i)}.
\label{eq:cs}
\end{equation}

By now, we can use the estimated \emph{counterfactual satisfaction} as the reward signal to update the RL policy by optimizing \myeq{L_rl}. The whole process is illustrated in the pre-learning and RL planning stage in \myfig{method}. 
We use the Adam optimizer \cite{adam} in learning the causal user model $\phi_M$, the policy $\pi_\theta$, and the state tracker. 

\smallskip
At last, by innovatively enhancing the offline RL framework with causal inference, we obtain a policy that can guarantee large user satisfaction by preventing filter bubbles, i.e., overexposing items.

\section{Experiments}
\label{sec:exp}
In this section, we conduct experiments to evaluate the IRS. 
% This is the RL evaluation stage in \myfig{method}.
We aim to investigate the following research questions:\\
\textbf{(RQ1)} How does CIRS perform compared with SOTA static recommendation methods and RL-based interactive recommendation policies?\\
\textbf{(RQ2)} How does CIRS perform in a limited number of interactive rounds?\\
\textbf{(RQ3)} How does CIRS perform in different environments with varying user tolerance of filter bubble?\\
\textbf{(RQ4)} What is the effect of the key parameters in CIRS?

\subsection{Experimental Setup}
We introduce the experimental settings with regards to the settings, environments, evaluation metrics, and state-of-the-art recommendation methods.

\subsubsection{Evaluation in the Interactive Recommendation Setting}
\label{seq:setting}
\begin{figure}[!t]
\centering
\includegraphics[width=1\linewidth]{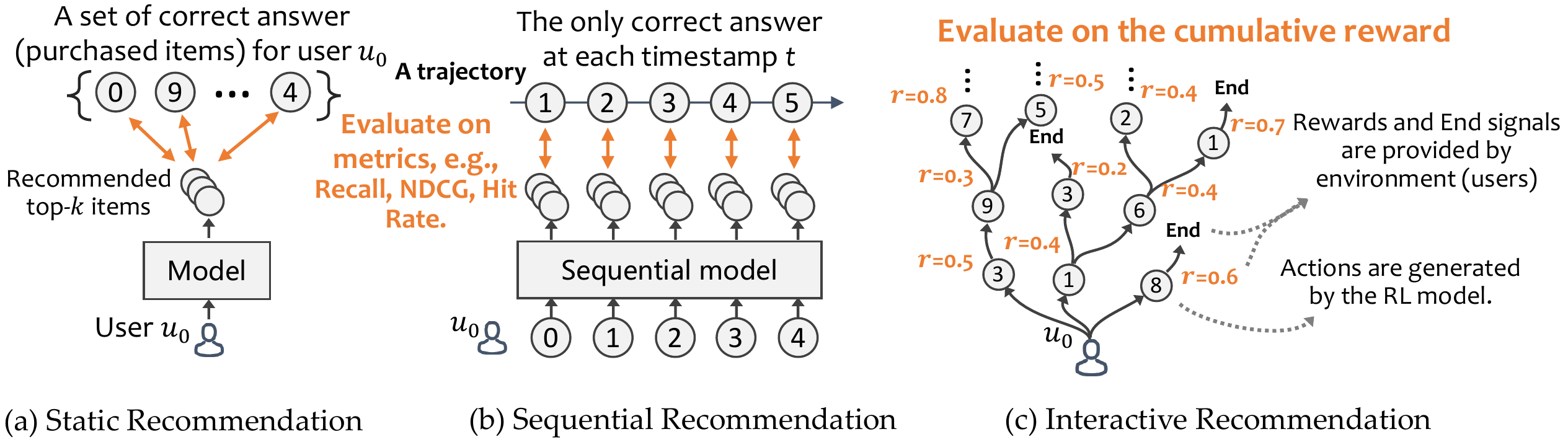}
% \vspace{-4mm}
\caption{Illustration of evaluation in three different recommendation settings.}
% \vspace{-4mm}
% cumulative satisfaction, interaction sequence length, and the reward per turn in
\label{fig:settings}
\end{figure}

We emphasize that we evaluate all methods in the interactive setting rather than traditional static or sequential settings. \myfig{settings} illustrates how static recommendation, traditional sequential recommendation, and interactive recommendation evaluate the models. Both the static and sequential recommendations use the philosophy of supervised learning, i.e., evaluating the top-$k$ results by comparing them with a set of ``correct'' answers in the test set and computing metrics such as Precision, Recall, NDCG, and Hit Rate. By contrast, interactive recommendation evaluates the results by accumulating the rewards along the interaction trajectories. There is no standard answer in interactive recommendation, which is interesting yet challenging. This setting requires offline data of high quality, which hampers the related research.

Studying filter bubbles in the interactive recommendation setting is necessary. Filter bubble is the phenomenon that occurs in real-world recommendation scenarios when the recommender overexposes similar content to a user. This means user satisfaction can change dynamically, hence it is inappropriate to study the filter bubble in traditional static or sequential settings where user preference in the test set is fixed. 

For now, interactive recommendation setting has not been widely investigated because it is hard to evaluate the model on the offline data. We overcome this problem by evaluating in the VirtualTaobao and KuaiEnv environments which are described in the following section.

\subsubsection{Recommendation Environments}
\label{seq:env}
Traditional recommendation datasets are too sparse to evaluate the interactive recommender systems.
We use two recommendation environments, VirtualTaobao\footnote{\url{https://github.com/eyounx/VirtualTaobao}} \cite{virtualtaobao} and KuaiEnv\footnote{\url{https://kuairec.com}}. The two environments can play the same role as the online real users. For the recommenders, an environment is like a black box as shown in the upper right corner of \myfig{method}.

% \begin{table}[!t]
% \tabcolsep=6pt
% \small
% \caption{Statistics of the KuaiEnv environment. The \emph{small matrix} is fully filled. The \emph{big matrix} is partially filled and contain all users and items in the \emph{small matrix}.}
% % \vspace{-2mm}
% \label{tab:kuai}
% \begin{tabular}{@{}lccccc@{}}
% \toprule
%                       & \textbf{\#Users} & \textbf{\#Videos} & \textbf{\#Attributes} & \textbf{Density} \\ \midrule
% \emph{small matrix}         &   1,411    & 3,327            &          31           & $    99.6\%$  \\                             
% \emph{big matrix}           &   7,176    & 10,728           &          31           & $    12.77\%$  \\                             
% \bottomrule
% \end{tabular}
% \vspace{-3mm}
% \end{table}

\smallskip
\textbf{VirtualTaobao} is a benchmark RL environment for recommendation. It is created by simulating the behaviors of real users on Taobao, one of the largest online retail platforms, via a multi-agent adversarial imitation learning (MAIL) approach. The simulated users can imitate the behavior of the real users and generate the same statistics as recorded on the Taobao platform. In the VirtualTaobao environment, a user is represented as an $88$-dimensional vector $\mathbf{e}_{u} \in \{0,1\}^{88}$, and a recommendation is represented as a $27$-dimensional vector $\mathbf{e}_{i} \in \mathbb{R}^{27}, \mathbf{0}\leq \mathbf{e}_{i} \leq\mathbf{1}$. When a model makes a recommendation $\mathbf{e}_{i}$, the environment will immediately return a reward signal representing user interest, i.e., a scalar $r\in \{0,1,\cdots, 10\}$.

It provides $100,000$ logged interactions for training the offline RL policy. Since the items are represented as the continuous vectors in VirtualTaobao, we use Euclidean distance to compute the distance between two items, i.e., $\operatorname{dist}(i,i_l)$ term in \myeq{ee} and \myeq{eestar}.

\smallskip
\textbf{KuaiEnv} is created by us on the KuaiRec dataset \cite{gao2022kuairec}. 
KuaiRec is a real-world dataset that contains a fully-observed user-item interaction matrix. The term ``fully-observed'' means that each user has viewed each video in the whole set and then left feedback. 
Therefore, unlike VirtualTaobao which simulates real users by training a model on Taobao data, i.e., the reward representing user preference is provided from a generative model, KuaiEnv uses real user historical feedback for each user-item pair, which can be more persuasive. We define the reward signal as the video watching ratio which is the ratio of viewing time to the total video length. Without loss of generality, we use this floating-point number to indicate users' intrinsic interest, i.e., we assume that the preference of a user keeps static and equals the logged ratings.
We use the fully-observed matrix, i.e., \emph{small matrix}, to evaluate the policy $\pi_\theta$. For pre-learning the user model $\phi_M$, we use the additional sparse user–video interactions in the \emph{big matrix}. In the RL planning stage, we learn the policy model $\pi_\theta$ by using the rewards provides by user model $\phi_M$ without levering offline data. For the details of the data, please refer to the KuaiRec dataset \cite{gao2022kuairec}.

% Briefly, we first select a set of users and videos on the Kuaishou App, then we alter the online recommendation rule to make sure each user has received the recommendation of every video. Hence, we obtain a fully filled user-item matrix, dubbed as the \emph{small matrix}. 
% It is used to evaluate the policy $\pi_\theta$. For pre-learning the user model $\phi_M$, we collect additional sparse user–video interactions in a \emph{big matrix}. We will release this dataset along with its detailed descriptions soon.
% Concretely, we first collected the users and videos labeled with ``high quality'' by the data analytics team of Kuaishou. Then we filtered a subset of them aiming to construct a fully filled matrix, i.e., each user has viewed all videos in this subset. For filling the inevitable missing values, we altered the online recommendation rule to insert these videos into the online recommendation streaming to make sure all of them were recommended to the user. 
% It took $15$ days for this exposure process, and finally we obtained a fully filled matrix, dubbed as \emph{small matrix}. 
% Since it contains all users' intrinsic interest over each video, we use the \emph{small matrix} to evaluate the policy $\pi_\theta$. For pre-learning the user model $\phi_M$, we collect additional sparse user–video interactions in a \emph{big matrix}.

% \mytable{kuai} illustrates the statistics of \emph{small matrix} and \emph{big matrix}. 
% The statistics of the KuaiRec dataset can be referred to \citet{gao2022kuairec}.
Each video in KuaiRec has at least one and no more than four categorical attributes, e.g., Food or Sports. Hence we use the Hamming distance for computing the term $\operatorname{dist}(i,i_l)$ in \myeq{ee} and \myeq{eestar}. 

\begin{figure*}[t!]
\centering
\includegraphics[width=0.8\linewidth]{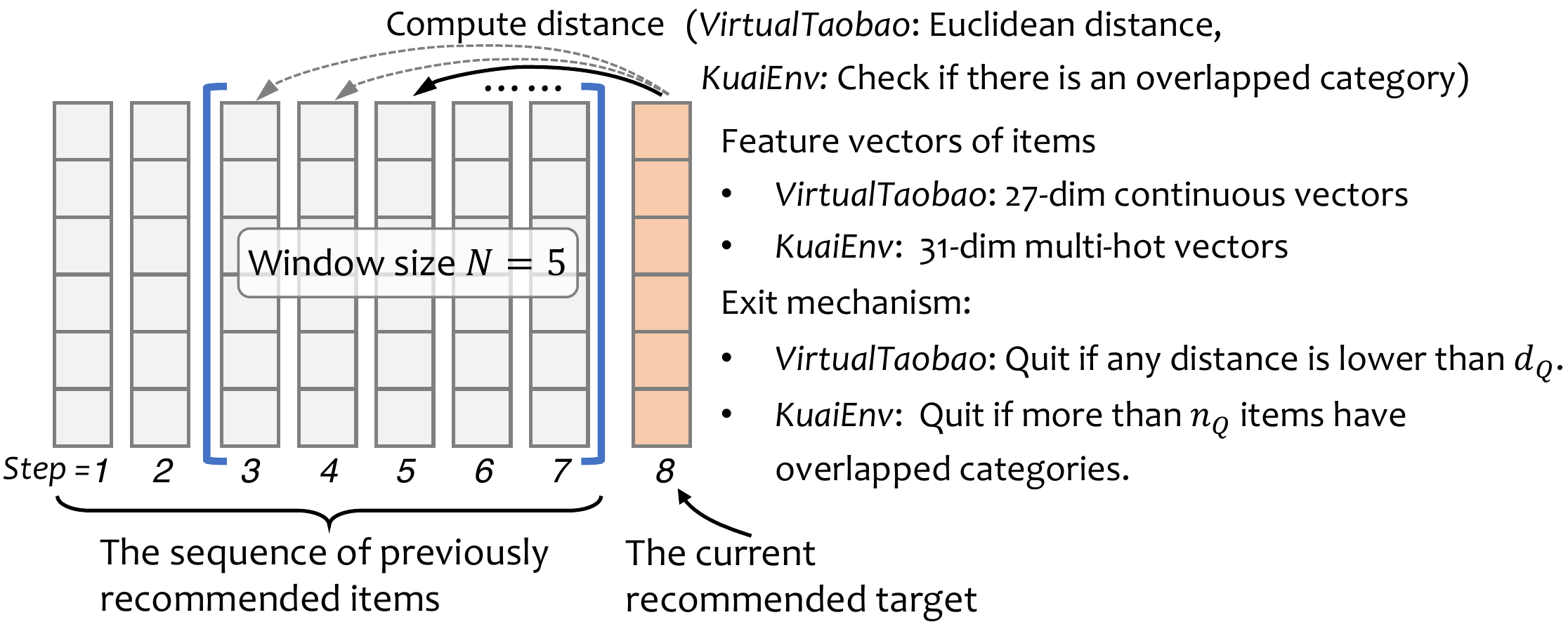}
% \begin{tabular*}{0.9\linewidth}{@{\extracolsep{\fill}}cc}
% \textbf{(a-b) Results with large interaction rounds} & (c-d) Results with limited interaction rounds.
% \end{tabular*}
% \vspace{-6mm}
\caption{Illustration of the exit mechanism.}
% \vspace{-3mm}
% cumulative satisfaction, interaction sequence length, and the reward per turn in
\label{fig:exit}
\end{figure*}

\smallskip
\textbf{Exit Mechanism.} 
By now, VirtualTaobao and KuaiEnv can provide users' intrinsic interest as the reward signal. However, they cannot reflect users' responses to the overexposure effect. To this end, we introduce the ``feel bored then quit'' mechanism in two environments to penalize filter bubbles in evaluation.
Usually, the environments will repeatedly interact with the recommender. VirtualTaobao has only a naive mechanism for ending the interaction by predicting the length of the interaction trajectory in advance. It is not in control and we alter it by considering the observations found in \mysec{verification}. 
The exit mechanism is illustrated in \myfig{exit}. Concretely, we compute the Euclidean distance between the recommended target and the most recent $N$ recommended items. If any of them is lower than the threshold $d_Q$, the environment will quit the interaction process as the real users can feel bored and quit under such monotonous recommendation. 
In KuaiEnv, similarly, for the most recent $N$ recommended items, if there are more than $n_Q$ items in the $N$ items have at least one attribute of the current recommended target, then the user in this environment ends the interaction process. Intuitively, a good recommender should avoid repeating highly similar items to prevent users from quitting early.

\subsubsection{Evaluation Metrics}
% We aim to pursue long-term success. Therefore, we compute the average cumulative satisfaction of $100$ sequences of the interaction between the environment and the recommender. To achieve a large cumulative satisfaction, the recommender policy needs to find a trade-off between pursuing a high single-round satisfaction and maintaining a longer interaction sequence.

% In the RL evaluation stage, we deploy the learned policy in the real environment (mimic environment/dynamic scenario) as described in \mysec{env} (as shown in \myfig{method}) to verify the effectiveness of our model. As in the actual scenario, the recommender system is always required to protect users' privacy, and users' profile information and interactions with the system are confidential. (For security reasons, they will not disclose their system composition, algorithms, etc.) We consider such a realistic black-box scenario, which means the recommender system is totally agnostic to the agent (as shown in \myfig{method}). 
We aim to evaluate the model performance with regard to cumulative satisfaction over the whole interaction trajectory $\mathcal{S}$, i.e., $\sum_{l=0}^{|\mathcal{S}|} r_{t+l}$,
% \begin{equation*}
%     \sum_{l=0}^{|\mathcal{S}|} r_{t+l},
% \end{equation*}
where $r_t$ is the reward signal returned by VirtualTaobao or KuaiEnv. Note that in this setting, user satisfaction is set as:
\begin{equation*}
\vspace{-1mm}
\text{satisfaction}=\left\{
\begin{array}{rcl}
\text{interest}, & & \text{if no filter bubble ever occurs}, \\
0, & & \text{otherwise}.
\end{array} \right.
% \vspace{-2mm}
\end{equation*}
I.e., if the recommendation does not trigger the exit mechanism, we can accumulate the rewards to represent intrinsic interest. But whenever an overexposed item triggers the exit mechanism, the interaction is interrupted and no reward can be added anymore. Thereby, the system cannot repeat to recommend the several high-quality items of the maximum confidence.
Intuitively, to pursue long-term success, the recommender policy must find a trade-off between pursuing a higher single-round satisfaction and maintaining a longer interaction sequence. 

We report the average cumulative satisfaction over $100$ interaction sequences. 

% By continuously interacting with the recommender system, users' immediate feedback (i.e., satisfaction over recommended items) is obtained and regarded as the reward signal guiding the subsequent optimization process. Since our goal is to achieve higher cumulative satisfaction, we record each interaction's average cumulative satisfaction of $100$ sequences. Besides, to pursue long-term success, the recommender policy must find a trade-off between pursuing a higher single-round satisfaction and maintaining a long interaction sequence.

\subsubsection{Baselines}
We use the commonly used static recommendation models plus straightforward policies as baselines. 
For KuaiEnv which has rich item features, four static recommendation baselines are:
\begin{itemize}
	\item \textbf{DeepFM} \cite{DeepFM}, which is a powerful factorization machine-based neural network containing wide and deep parts to extract knowledge from low- and high-order feature interactions. It serves as a strong backbone in the recommender framework in many companies.
	\item \textbf{IPS} \cite{swaminathan2015counterfactual} is a well-known statistical technique adjusting the target distribution by re-weighting each sample in the collected data. In recommendation, it is widely used for modeling the probability of observation in order to remove the exposure bias or selection bias in the collected data. It is easy to implement and suffers from the high variance issue \cite{10.5555/3045390.3045616}.
	\item \textbf{PD} (Popularity-bias Deconfounding) \cite{zhangyang} is a causal inference-based method that models item popularity as a confounder introducing spurious correlations between exposed items and user preference. By explicitly modeling popularity, PD can remove the popularity bias in the final recommendation stage.
	\item \textbf{DICE} \cite{DICE} tries to disentangle popularity and user interest by separately modeling them in the so-called causal embedding. Therefore, item popularity or other unwanted factors can be removed in the recommendation stage.
\end{itemize}
It should be noticed that:
(1) IPS, PD, and DICE are techniques used for debiasing, and we implement their backbone network as DeepFM. 
(2) All these methods are deterministic/static models, in which researchers usually take the items with top-$1$ or top-$k$ highest predicted scores as the final recommendations. In our interactive setting, this manner will incur overexposure immediately. Therefore, we make the recommendation by sampling from the final logits with a Softmax layer. 

We also implement basic policies in KuaiEnv:
\begin{itemize}
	\item \textbf{Random}, which recommends random items completely.
	\item \textbf{$\mathbf{\epsilon}$-greedy}, which outputs a random result with probability $\epsilon$ and uses the results from the DeepFM model with probability $1-\epsilon$.
	\item \textbf{UCB} maintains an upper confidence bound for each item and follows the principle of optimism in the face of uncertainty. It means if we are uncertain about an action, we should give it a try. UCB can balance the exploration and exploitation in decision-making process.
\end{itemize}

% the static recommendation modules in the user model are implemented as the powerful baseline \textbf{DeepFM} \cite{DeepFM} and three representative debiasing methods: \textbf{IPS} \cite{swaminathan2015counterfactual}, \textbf{PD} \cite{zhangyang}, and \textbf{DICE} \cite{DICE}. While IPS is a classic method used in a lot of works 

% We implement the debiasing modules of the three methods on the basic DeepFM model. Since these static methods do not have a strategy for varying the recommendation results, we add a Softmax layer to their predicting results and randomly sample items from the distribution. For comparing with the other policies, we add several effective policies including \textbf{Random}, \textbf{$\mathbf{\epsilon}$-greedy}, \textbf{UCB} \cite{UCB} to the results of DeepFM. 
For VirtualTaobao, since the users and items are given by feature vectors, we can only implement a multilayer perceptron (\textbf{MLP}) with sampling on the Softmax results and the $\epsilon$-greedy strategy. 
Besides, we implement the powerful RL baseline, PPO \cite{ppo}, on the same offline RL framework, i.e., it is learned by interacting with the user model but without the causal inference module. For comparison, we denoted this baseline as $\textbf{CIRS w/o CI}$. 

Note that PPO is one component in the model-based offline RL and it can be replaced by other RL models (e.g., DQN \cite{dqn}, Actor-Critic \cite{a2c}, or DDPG \cite{ddpg}). $\textbf{CIRS w/o CI}$ can be deemed as a framework that summarizes model-based offline RL methods \cite{jinhuang22,Keeping-recsys}. For example, the difference with \citet{Keeping-recsys} is that they used MF as the user model while we use DeepFM; the difference with \citet{jinhuang22} is that they investigated a lot of sequential models as state trackers while we use a Transformer-based model.

\subsection{Overall Performance Comparison}
\label{sec:large_round}
We evaluate the proposed CIRS and baselines in two environments. We use grid search to tune the optimal parameters for all methods. For example, in VirtualTaobao, the key parameter $\tau^*$ is searched in $\{0.001, 0.005, 0.01, 0.1, 0.5, 1.0, 5.0\}$ and $\tau$ is searched in $\{0.001, 0.005, 0.01, 0.05, 0.1, 0.5, 1.0\}$. The results are illustrated in \myfig{tau}. For more implementation details, please visit the instruction via this Github link\footnote{\url{https://github.com/chongminggao/CIRS-codes/tree/main/reproduce_results_of_our_paper}}.

For general comparison, we do not limit the length of the interaction and set the max round to be large enough (but feasible to implement). We set the max round to be $50$ and $100$ for VirtualTaobao and KuaiEnv, respectively. For the parameters in the environment setting, we set the window size, i.e., the number of the most recent recommendations, to be $N=5$ and the exit threshold to be $d_Q = 3.0$ for VirtualTaobao and $N=1$, $n_Q=1$ for KuaiEnv. 

\begin{figure*}[t!]
\centering
\includegraphics[width=1\linewidth]{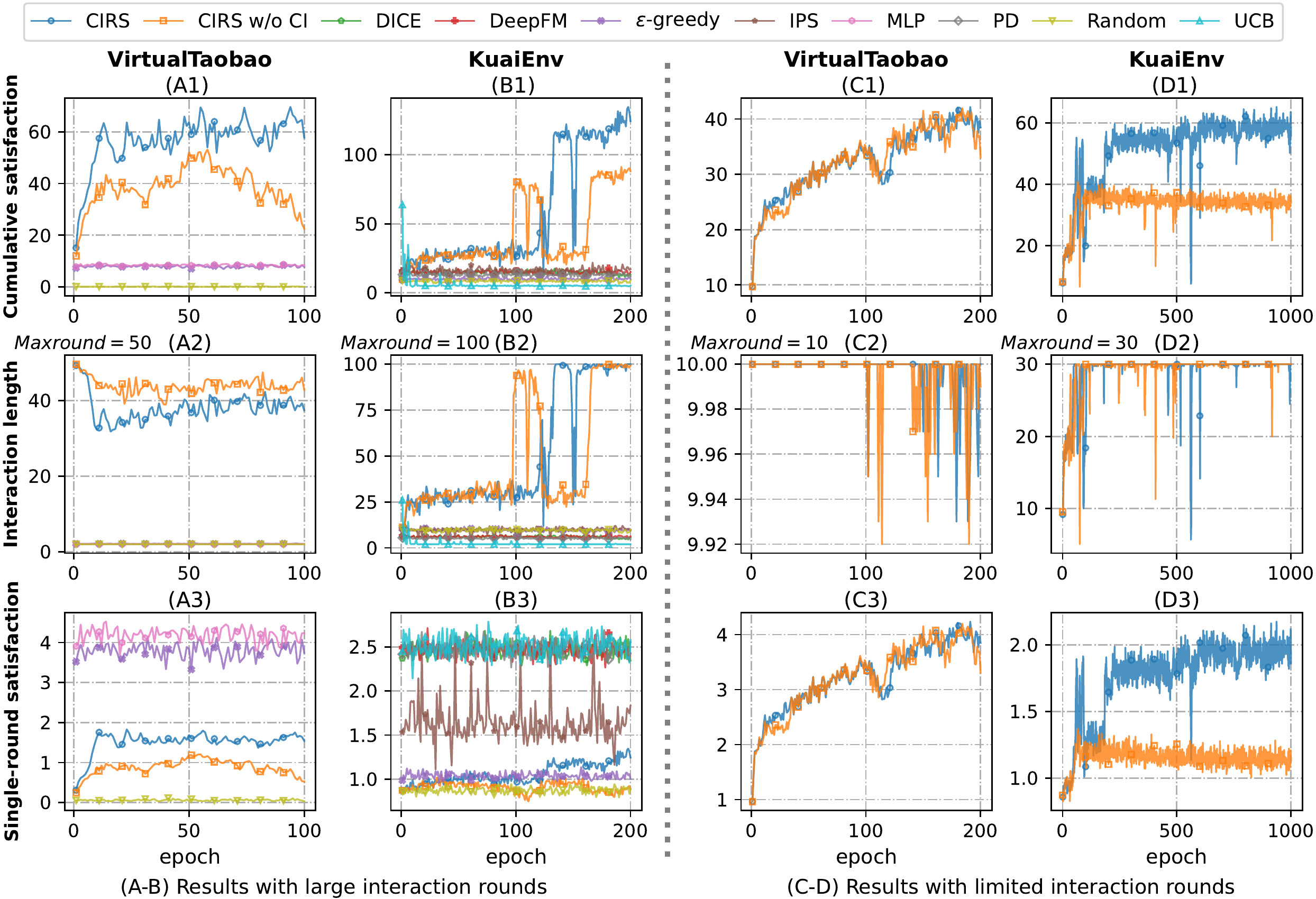}
% \begin{tabular*}{0.9\linewidth}{@{\extracolsep{\fill}}cc}
% \textbf{(a-b) Results with large interaction rounds} & (c-d) Results with limited interaction rounds.
% \end{tabular*}
% \vspace{-6mm}
\caption{Results of all methods with large interaction rounds (\mysec{large_round}) and limited interaction rounds (\mysec{limited_round}).}
% \vspace{-3mm}
% cumulative satisfaction, interaction sequence length, and the reward per turn in
\label{fig:main_result}
\end{figure*}

The results are shown in \myfig{main_result} (A-B). The first row shows the cumulative satisfaction, which is the global metric to evaluate the recommender systems. The second row and third rows show the details of the user satisfaction, i.e., the length of the interaction trajectory and the single-round satisfaction, respectively. From (A1) and (B1), we can see the proposed CIRS achieves the maximal average cumulative satisfaction after several epochs in both VirtualTaobao and KuaiEnv.

In the first few epochs in VirtualTaobao, the performances of both CIRS and CIRS w/o CI improve because the RL policy gradually finds the correct user preference so the satisfaction in each round increases (A3). Interestingly, the increase in the single-round satisfaction compromises the length of trajectory at the beginning (A2). Later, the length gradually becomes stable, and the policy of CIRS eventually finds a balance point between length and single-round satisfaction, thus achieving the maximal cumulative satisfaction. However, without the causal inference module, i.e., as shown by CIRS w/o CI, the policy becomes unstable and the performance degenerates with the epoch increases. 
This phenomenon demonstrates the effectiveness of causal inference in capturing the overexposure effect and thus avoiding repeatedly recommending items.

In KuaiEnv, CIRS also achieves the largest cumulative satisfaction (B1) after enough epochs. 
Unlike in VirtualTaobao, the performance increases mainly because of the increase in the interaction length. As shown in (B2), the interaction lengths of both CIRS and CIRS w.o. CI increases to the maximum length of $100$ after about $160$ epochs. In addition, the cumulative satisfaction of CIRS further increases at about $180$ epochs in (B1), which is due to the possibility of further improving the single-round performance (B3).
For both VirtualTaobao and KuaiEnv, CIRS beats the counterpart method CIRS w/o CI, which demonstrates the effectiveness of the causal module in CIRS. 

For other baselines, we can see that all other methods except Random can achieve better single-round performance (A3 and B3). However, their recommendation results are too limited and narrow even with the randomness introduced by the basic policies (i.e., Random sampling, Softmax-based sampling, and $\epsilon$-greedy). Note that in VirtualTaobao, even the random sampling cannot bring a longer interaction sequence because of the curse of dimensionality: The action space has $88$ dimensions, therefore, the Euclidean distance of any two random points becomes statistically indiscriminate. The result of IPS fluctuates intensely in terms of the single-round performance (B3), which is due to the widely discussed high variance issue \cite{swaminathan2015counterfactual}. 
The interaction lengths of $\epsilon$-greedy and IPS are longer than other methods (i.e., DICE, PD, and DeepFM) in (B2). This is because the two methods have the ability to explore the item space during the whole interaction process. Compared with these two naive methods, UCB is a policy that can automatically balance exploration and exploitation, it has the best performance at the beginning. However, after several epochs of exploration, the policy enhances its belief in certain items and thus leads to getting stuck in the filter bubble. Therefore, UCB ends up with the lowest interaction length as shown in (B2) but with the maximum single-round satisfaction in (B3). 

To conclude, except for the deep RL policy-based methods (i.e., CIRS and CIRS w/o CI), static recommendation models with heuristic policies (i.e., Softmax-based sampling, and $\epsilon$-greedy, and UCB) cannot overcome the overexposure effect, thus lead to the filter bubbles and result in low user satisfaction. Furthermore, by comparing CIRS with CIRS w/o CI, we show the effectiveness of causal inference in the offline RL framework.

\subsection{Results with Limited Interaction Rounds}
\label{sec:limited_round}
In real-world recommendation scenarios, users have limited energy and will not spend too many rounds interacting with the recommender. Therefore, we limit the max round to $10$ and $30$ in VirtualTaobao and KuaiEnv, respectively. 
We aim to investigate whether the policy can exploit to improve the single-round satisfaction under such a situation.
We alter the exit threshold $d_Q = 1.0$ in VirtualTaobao for better demonstration. 

From the results in \myfig{main_result} (C-D), we can see that CIRS outperforms CIRS w/o CI in KuaiEnv (D1 and D3), and produces a similar performance in VirtualTaobao (C1 and C3). In VirtualTaobao, both the two policies achieve the same level of single-round performance (greater than $4.0$) after approximately $150$ epochs. And the performance is even similar to the static methods (C3 and A3). In KuaiEnv, both the two policies achieve higher single-round performances (D3) compared with that in the former setting (B3). Especially, CIRS has a great improvement in single-round performance (D3 and B3), which means that it is suitable for real-world interactive recommendation scenarios with limited interaction rounds. Actually, in (B1 and B3), we can also see that the performance of CIRS continues to increase at $epoch=200$. This means CIRS has potential even under huge rounds, given enough training time. The results in (C3 and D3) show that the knowledge, i.e., the correct user preference can be distilled from the causal user model to the RL policy. This further demonstrates the effectiveness of our model-based RL framework.
Again, in this setting, we conclude that by integrating with causal inference, we let CIRS outperform its counterpart. 

\begin{figure}[!t]
\centering
\includegraphics[width=.7\linewidth]{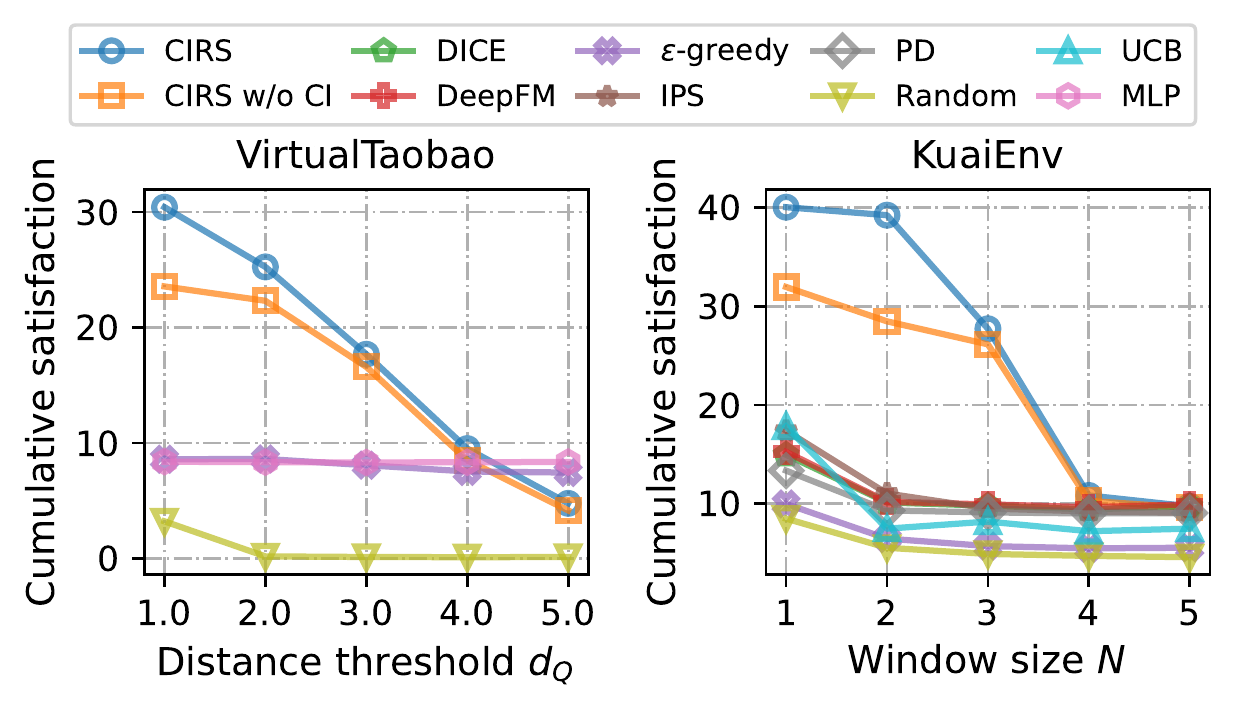}
% \vspace{-4mm}
\caption{Results under different user sensitivity}
% \vspace{-4mm}
% cumulative satisfaction, interaction sequence length, and the reward per turn in
\label{fig:leave}
\end{figure}

\subsection{Results with Different User Sensitivity}
\label{sec:vary_env}
To validate the generality of CRS, we vary the parameters: the distance threshold $d_Q$ and window size $N$ to illustrate their effects in VirtualTaobao and KuaiEnv, respectively. A large $d_Q$ or $N$ means that users get more sensitive to filter bubbles and become easier to quit the interaction. The results in \myfig{leave} show that CIRS outperforms all baseline methods when users are less sensitive, i.e., small $d_Q$ in VirtualTaobao and small $N$ in KuaiEnv. CIRS obtains the best cumulative satisfaction because it can avoid repeating recommending highly similar items and thus can maintain a long interaction length.

However, the performance of CIRS inevitably decreases when users become more sensitive, though it can still beat its counterpart CIRS w/o CI. When $d_Q \geq 4$ or $N \geq 4$, CIRS and CIRS w/o CI can only achieve the same or even worse performance compared with other baselines. This means that facing extremely picky users, even the model enhanced by causal inference cannot alleviate the dissatisfaction caused by overexposing any item. When $d_Q = 5$ or $N = 5$, the two RL-based models even cannot beat DeepFM or MLP which are served as the teacher models in the two model-based RL methods. This is because the RL-based methods do not have the opportunity to conduct their explore-exploit philosophy when the user is too picky.

Meanwhile, other baselines have similar performance under different user sensitivity --- the recommendations make users feel bored and quit even though the user is more tolerant to filter bubbles (i.e., less sensitive to item overexposure). This also demonstrates they are not suitable for addressing the filter bubble issue.

\begin{figure}[!t]
%   \vspace{-2mm}
    \tabcolsep=-1pt
    \centering
    \begin{tabular}{cc}
    \includegraphics[width=0.45\linewidth]{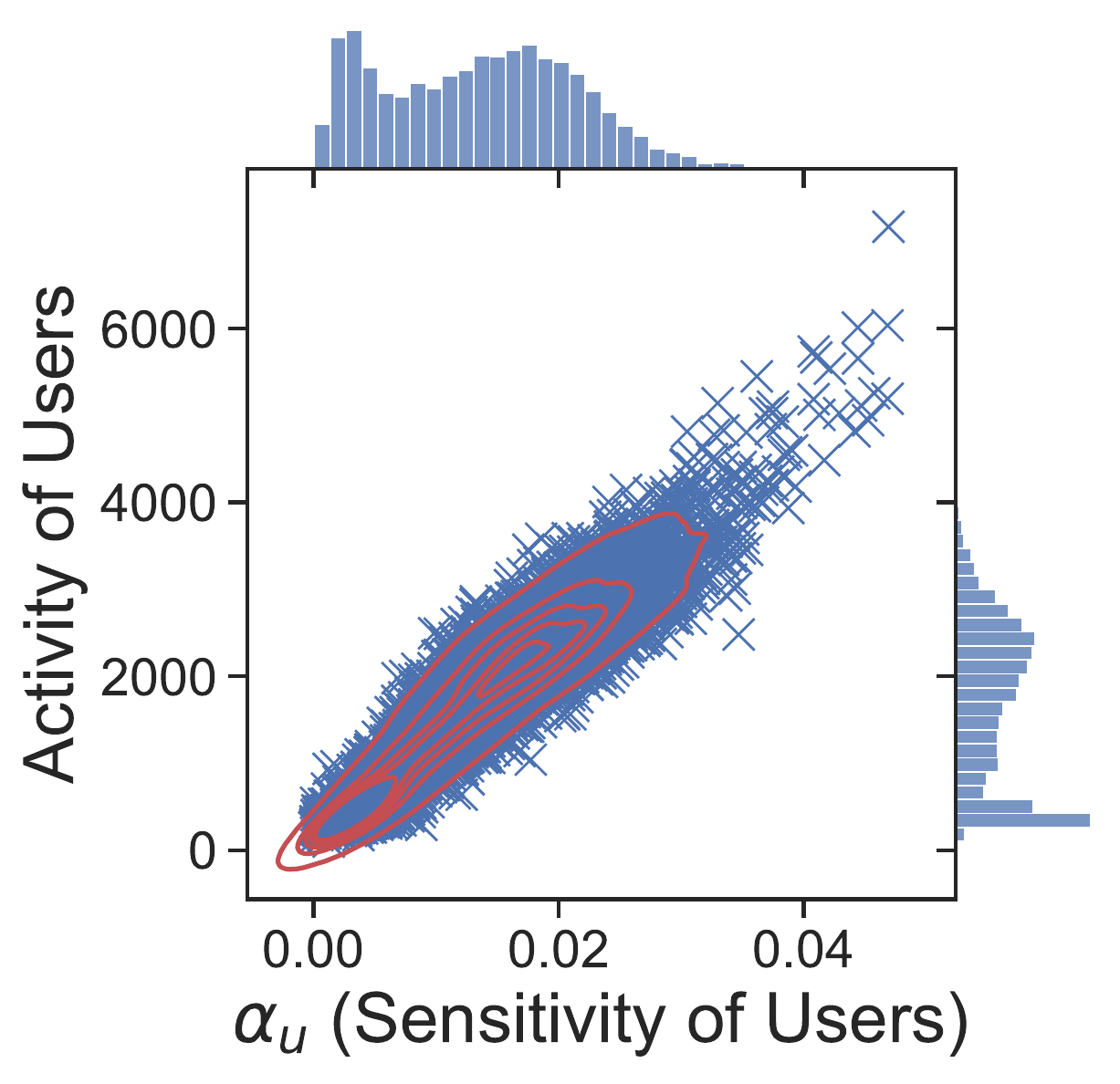} & \includegraphics[width=0.45\linewidth]{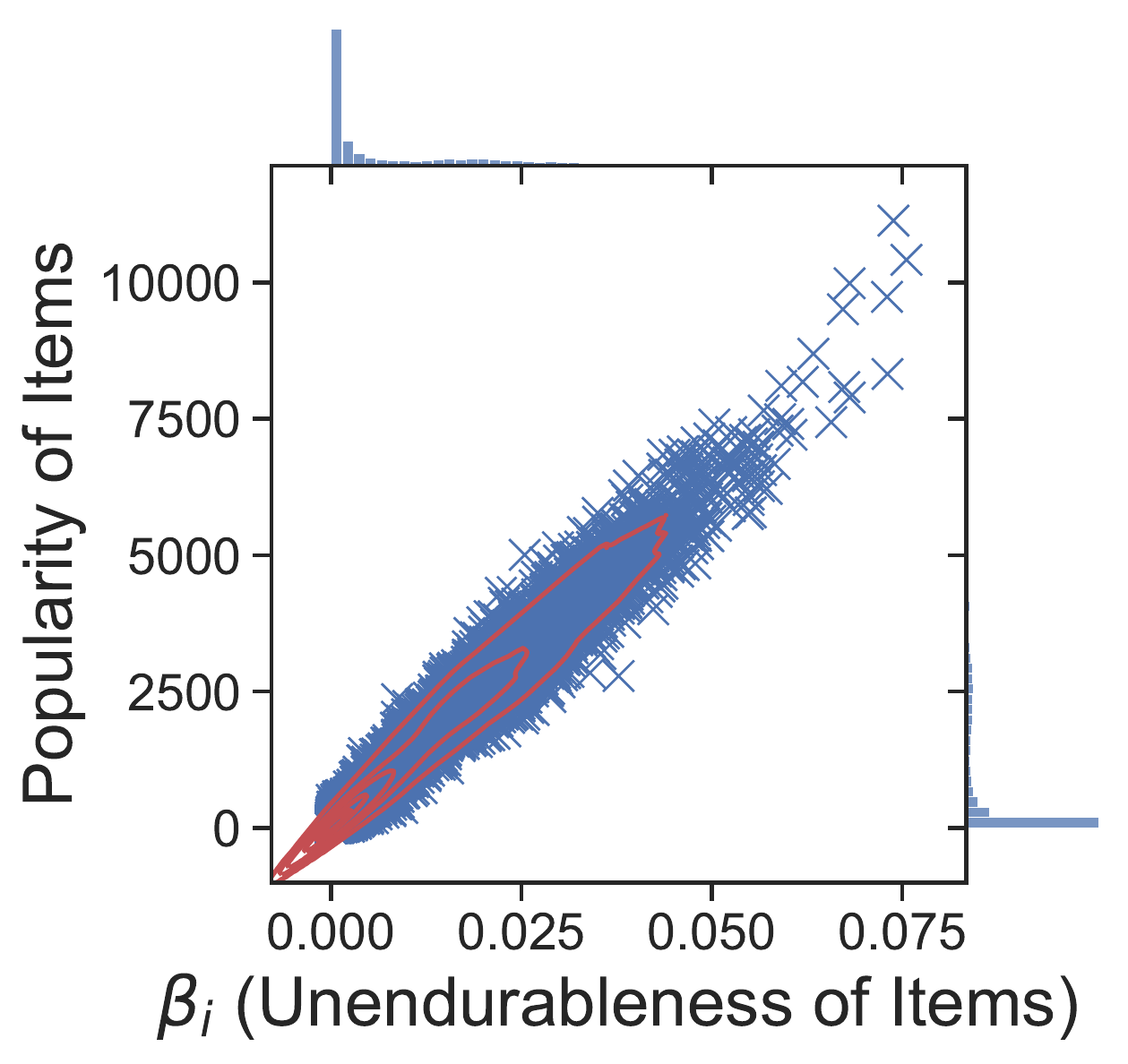}
    \end{tabular}
    % \vspace{-4mm}
    \caption {Relationship between the learned $\alpha_u, \beta_i$ and the data statistics (user activity and item popularity).}
    \label{fig:ab}
\end{figure}

\begin{figure}[!t]
\centering
% \vspace{-4mm}
\includegraphics[width=0.8\linewidth]{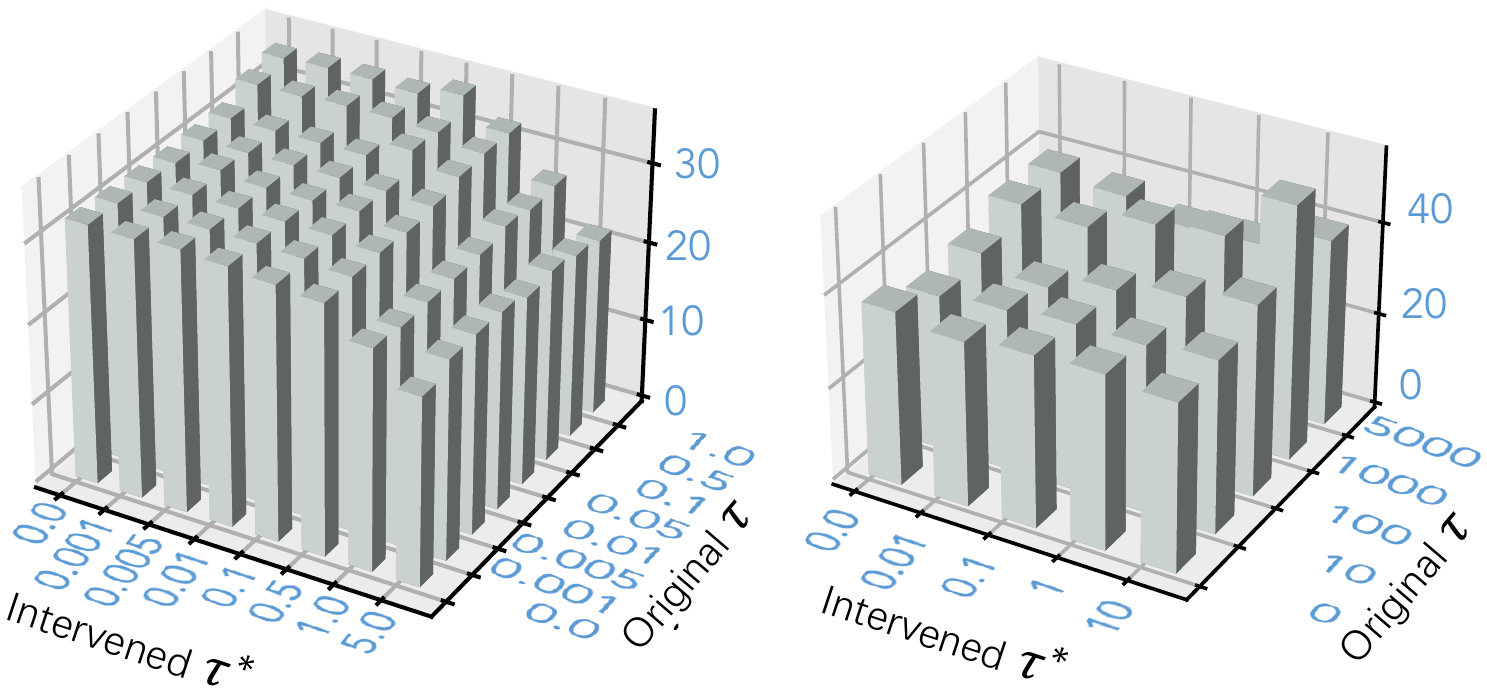}
\begin{tabular*}{0.5\linewidth}{@{\extracolsep{\fill}}cc}
(a) VirtualTaobao & (b) KuaiEnv
\end{tabular*}
% \vspace{-3mm}
\caption{Cumulative satisfaction with different $\tau$-$\tau^*$ pairs.}
% \vspace{-4mm}
% cumulative satisfaction, interaction sequence length, and the reward per turn in
\label{fig:tau}
\end{figure}

\subsection{Effect of Key Parameters}
\label{sec:key_p}
We investigate the effect of the key parameters of CIRS in KuaiEnv. We compare the learned $\alpha_u$ and $\beta_i$ in the user model with two statistics of data, i.e., the activity of the users and the popularity of items, which are derived by summing the rows and columns of the \emph{big matrix}. We show the results in \myfig{ab}. The results are intuitive to understand: user sensitivity, i.e., $\alpha_u$, is proportional to the user activity, i.e., an active user is easier to get bored when viewing overexposed videos because he/she may have seen many similar content before. Similarly, item unendurableness, i.e., $\beta_i$, is proportional to the item popularity, i.e., popular videos are less endurable when they are overexposed. This also explains why popular items become outdated quickly.

Furthermore, we investigate the effect of different combinations of $\tau$ (in \myeq{ee}) and $\tau^*$ (in \myeq{eestar}) on the cumulative satisfaction. CIRS with $\tau=0, \tau^*=0$ degenerates to CIRS w/o CI since both $e_t(u,i)$ and $e_t^*(u,i)$ become $0$, i.e., the modeling of user satisfaction will not take into account the overexposure effect. The results in \myfig{tau} demonstrate that suitable $\tau$-$\tau^*$ pairs indeed improve the performance compared to CIRS w/o CI. Note that the orders of magnitude of the $\tau$-axis and $\tau^*$-axis differ greatly in KuaiEnv, because the unit of time in \myeq{ee} and \myeq{eestar} are different. The former uses the second(s) in log data and the latter uses the step(s) in the RL planning and evaluation stages.

\section{Conclusion and Discussion}
This work studies filter bubbles in the interactive recommendation setting. Different from the static and sequential recommendation settings which use the philosophy of supervised learning, the interactive setting evaluates the RL-based policies by accumulating the rewards along the interaction trajectories.

The interactive recommendation setting provides a practical way to track and estimate the filter bubble, which is the phenomenon that occurs in real-world recommendation scenarios when the recommender overexposes similar content to a user. We conduct field studies on music and video recommendation datasets and show that user satisfaction will drop with the increasing of similar content, which spurs us to remove filter bubbles in recommender systems.

We propose a counterfactual interactive recommender system (CIRS), leveraging causal inference in offline RL to deduce users' varying satisfaction. 
CIRS utilizes a causal user model that can disentangle the intrinsic user interest from the overexposure effect of items. The causal user model provides unbiased \emph{counterfactual rewards} for learning the RL policy. 
% The field studies on Kuaishou App verify that overexposing items indeed hurts the user experience, 
To conduct evaluations, we innovatively create a faithful RL environment, KuaiEnv, based on a real-world fully-observed user rating dataset. 
Extensive experiments demonstrate that the proposed method can burst filter bubbles and increase users' cumulative satisfaction.
The experiments show that CIRS can obtain optimal cumulative satisfaction by finding the trade-off between pursuing a high single-round satisfaction and maintaining a long-lasting interaction.

Our work has several noteworthy contributions. The most important one is that we demonstrate the right way to evaluate RL-based methods in the interactive recommendation setting, i.e., evaluating the decision-makers (i.e., recommendation policies) by the cumulative reward. In reality, real users do not have any standard answers in their minds when they use recommender systems; and the companies care whether the model can make users satisfied in the long term. Therefore, the interactive recommendation setting can well describe real-world recommendation scenarios.

However, many previous works still evaluate the RL-based methods via the static or sequential settings, i.e., evaluating the top-k results by comparing them with a set of “correct” answers in the test set and computing metrics such as Precision, Recall, NDCG, and Hit Rate \cite{10.1145/3488560.3498471,Xinxin}. We understand why they chose to evaluate that way: the evaluation of RL is notoriously hard on offline data. To overcome this problem, we create the KuaiEnv environment in which each user's preference towards all items is known. With this environment, researchers can conduct faithful evaluation without having to synthesize user preference in simulated user-item matrices \cite{minmin_topK,Keeping-recsys,jinhuang22}.

By modeling and alleviating filter bubble issues in the interactive recommendation setting, we demonstrate the potential research directions and possible solutions in the recommendation community.
% In the future, it is interesting to explore other types of biases in interactive recommendation. This work illustrates filter bubble would evolve during the process of user-system interaction. We believe other biases may also have this nature. It will be valuable to transfer the experience of this work to tackle other biases in a dynamic environment. The proposed causality-enhanced IRS framework could be adapted for other biases. 
In the future, we believe that the interactive recommendation will draw a lot of research attention. It is interesting to explore other types of biases in this setting. Combining causal inference and reinforcement learning is promising since causal inference can provide an explicit guide to optimize models and thus introduce explainability in RL \cite{tutorialCausalRL,madumal2020explainable}.

\section*{Acknowledgements}
This work is supported by the National Key Research and Development Program of China (2021ZD0111802), the National Natural Science Foundation of China (61972372, U19A2079, 62121002), and the CCCD Key Lab of Ministry of Culture and Tourism.

\bibliographystyle{ACM-Reference-Format}
\bibliography{Counterfactual-IRS}

\end{document}